# Nonphotosynthetic Pigments as Potential Biosignatures

Edward W. Schwieterman,[1,2,3] Charles S. Cockell,[4,5] and Victoria S. Meadows[1,2,3]

## Abstract

Previous work on possible surface reflectance biosignatures for Earth-like planets has typically focused on analogues to spectral features produced by photosynthetic organisms on Earth, such as the vegetation red edge. Although oxygenic photosynthesis, facilitated by pigments evolved to capture photons, is the dominant metabolism on our planet, pigmentation has evolved for multiple purposes to adapt organisms to their environment. We present an interdisciplinary study of the diversity and detectability of nonphotosynthetic pigments as biosignatures, which includes a description of environments that host nonphotosynthetic biologically pigmented surfaces, and a lab-based experimental analysis of the spectral and broadband color diversity of pigmented organisms on Earth. We test the utility of broadband color to distinguish between Earth-like planets with significant coverage of nonphotosynthetic pigments and those with photosynthetic or nonbiological surfaces, using both 1-D and 3-D spectral models. We demonstrate that, given sufficient surface coverage, nonphotosynthetic pigments could significantly impact the disk-averaged spectrum of a planet. However, we find that due to the possible diversity of organisms and environments, and the confounding effects of the atmosphere and clouds, determination of substantial coverage by biologically produced pigments would be difficult with broadband colors alone and would likely require spectrally resolved data. Key Words: Biosignatures—Exoplanets—Halophiles—Pigmentation—Reflectance spectroscopy—Spectral models. Astrobiology 15, 341–361.

## 1. Introduction

THE SEARCH FOR LIFE beyond Earth requires an understanding of the measurable ways life can impact its environment. These measurable impacts are biosignatures: "an object, substance, and/or pattern whose origin specifically requires a biological agent" (Des Marais *et al.*, 2008). To be considered a useful biosignature, this evidence must be clearly discernable from the results of abiotic processes. Three main classes of remotely detectable biosignatures have been proposed. First, there are spectral absorption features produced by biosignature gases. These are gases that are produced by metabolic processes, and their production by abiotic processes, such as volcanic outgassing or photochemistry, can be reasonably excluded. The standard often applied is a requirement for a chemical disequilibrium that is most likely of biological origin (Lovelock, 1965; Hitchcock and Lovelock, 1967; Sagan *et al.*, 1993). Gaseous oxygen

($O_2$) and its photochemical by-product ozone ($O_3$) are examples of gaseous biosignatures that are robust to many false-positive scenarios, although recent work suggests that significant abiotic production of these gases may occur via photochemistry for stars with sufficiently high far-UV irradiances (*e.g.*, Domagal-Goldman *et al.*, 2014; Tian *et al.*, 2014) or via massive runaway-greenhouse-mediated water loss during the pre-main-sequence phase of low-mass stars (Luger and Barnes, 2015). Oxygen is considered a more robust biosignature in the simultaneous presence of a reduced gas like methane ($CH_4$). For a review of potential biosignature gases see, for example, the work of Seager *et al.* (2012). The second class of remotely detectable biosignatures consists of reflectance signatures from pigmented organisms, especially land vegetation. Reflectance is the fraction of incident light that is reflected from a surface, which will vary with wavelength depending on the optical properties of the surface. The most commonly explored spectral feature in this category is the vegetation red edge (VRE) of oxygenic photosynthesizers such as land-based plants, which produces a sharp increase in reflectance from the visible to the near-infrared portions of the spectrum (Sagan *et al.*, 1993; Seager *et al.*, 2005). Finally, there are temporal biosignatures. These are time-dependent changes in measurable quantities such as gas concentrations, planetary albedos, or broadband colors (the ratios of intensities measured between two or more filter bands) that are produced as a consequence of changes in biological activity (Meadows, 2006). For example, seasonal periodicities observed in $CO_2$ and $CH_4$ concentrations on Earth are correlated with land-based respiration (Keeling, 1960; Rasmussen and Khalil, 1981; Khalil and Rasmussen, 1983; Keeling *et al.*, 1996), and seasonal variation in pigmentation can be observed in land-based vegetation. The focus of this paper is on reflectance biosignatures with the acknowledgement that there may also be a seasonal or time-dependent component.

[1]University of Washington Astronomy Department, Seattle, Washington, USA.
[2]NAI Virtual Planetary Laboratory, Seattle, Washington, USA.
[3]University of Washington Astrobiology Program, Seattle, Washington, USA.
[4]University of Edinburgh School of Physics and Astronomy, Edinburgh, UK.
[5]UK Centre for Astrobiology, Edinburgh, UK.

The most commonly considered biosignatures are products of photosynthesis or are phenomena otherwise associated with photosynthetic organisms. Photosynthesis is a method of primary production (turning $CO_2$ to biomass) that uses photons from the Sun (or in general, a host star or stars) as a source of energy, and a reductant, such as $H_2$, $H_2S$, $Fe^{2+}$, or $H_2O$, as a source of electrons (Hohmann-Marriott and Blankenship, 2012). Oxygenic photosynthesis, which uses $H_2O$ as a reductant and generates $O_2$ as a waste product, is by far the most productive metabolism on modern Earth (Kiang *et al.*, 2007a). It is often reasoned that the organisms with the most productive metabolism will be the most plentiful and therefore will generate the most detectable signatures, and those signatures will be related to their primary metabolism.

However, oxygenic photosynthesis may not have always dominated the detectable biosphere in the past. Geological evidence suggests that biofilms or microbial mats of anoxygenic photosynthesizers may have generated the most prevalent surface signatures of life for several hundred million years before the development of oxygenic photosynthesis. This is inferred from the gap between the earliest stromatolite fossils at 3.5 Ga (Schopf, 1993) and the earliest undisputed geochemical evidence for oxygenic photosynthesis at 2.7 Ga (Buick, 2008).

There are currently no known remotely detectable gaseous biosignatures strictly associated with anoxygenic photosynthesis, and there are also many chemosynthetic metabolisms that do not produce waste gases in chemical disequilibrium with their environments (Des Marais *et al.*, 2002). In the absence of gaseous biosignatures, surface reflectance features would be the only possibly detectable biosignatures. Anoxygenic photosynthesizers and chemosynthetic pigment-bearing species would have generated their own surface reflectance biosignatures by reflecting more strongly at wavelengths where their pigment absorption was least efficient.

Near-future spaced-based telescopes may have the ability to directly image Earth-sized planets and search for surface reflectance biosignatures (Des Marais *et al.*, 2002; Cockell *et al.*, 2009; Levine *et al.*, 2009; Postman *et al.*, 2010; Seager *et al.*, 2014; Stapelfeldt *et al.*, 2014). To support these missions, it is important to better understand the possible diversity of planetary biosignatures, including surface biosignatures. Several authors have investigated the effects of photosynthetic vege-tation or microbial mats on the disk-averaged spectrum of Earth or Earth-like planets (Seager *et al.*, 2005; Montanes-Rodriguez *et al.*, 2006; Tinetti *et al.*, 2006b, 2006c; Sanromá, *et al.*, 2013), and Sanromá *et al.* (2014) investigated the spectrum and temporally varying broadband colors of an Earth substantially covered with anoxygenic purple bacteria. Hegde and Kaltenegger (2013) explored whether broadband filter photometry can serve as a first step to characterizing the surfaces of Earth-like exoplanets, and argued that visible spectrum broadband colors could be used to identify certain environments or significant coverage by specific types of extremophiles, lichens, or bacterial mats. However, this initial study did not consider the effects on the resulting broadband colors due to reflected radiation transmitted through the atmosphere.

As we will argue here, nonphotosynthetic pigments could provide alternative biosignatures to those generated by photosynthesis. In particular, this may be the case for highly productive chemosynthetic biospheres that have evolved pigmentation to cope with extreme environments, or for photosynthetic biospheres where the spectral reflectance is dominated by a nonphotosynthetic pigment. In support of this argument, we detail below a range of environments on Earth where nonphotosynthetic pigments dominate the spectral reflectance. While our focus is on applications to remotely detectable surface biosignatures on exoplanets, pigmented organisms may also be searched for in limited environments within our own solar system such as in the martian subsurface or Europa's ocean (Dalton *et al.*, 2003).

There is empirical evidence from several environments on modern Earth that, even in the presence of photosynthetic primary producers, the dominant reflectance biosignature can be from biologically produced pigments that developed to provide functions other than light capture for photosynthesis. For example, halophilic archaea such as *Halobacterium salinarum* and bacteria such as *Salinbacter ruber* dominate the spectral reflectance of hypersaline lakes and saltern crystallizer ponds with their nonphotosynthetic pigments (DasSarma, 2006; Oren, 2009, 2013). The pink coloration of the northern portion of the Great Salt Lake, Utah, USA, is visible in photographs from the International Space Station. Owens Lake, California, USA, and the saltern crystallizer ponds of San Francisco, USA, are further examples of the macroscopic coloration effect from these pigmented organisms (see Fig. 1a, 1b, and 1c). Other environments where pigmented halophiles dominate the spectral reflectance include Lake Hillier in Australia, the Sivash in Ukraine/Russia, and Lake Retba in Senegal. These halophilic organisms contain substantial amounts of carotenoids such as bacterioruberin, resulting in red, pink, or orange coloration to water above threshold salinities. Although the photosynthetic primary producers in hypersaline environments are often green algae like *Dunaliella salina,* which also contain carotenoid pigments, Oren *et al.* (1992) and Oren and Dubinsky (1994) found that the visible coloration of hypersaline lakes is dominated by the (smaller but more abundant) halophilic archaea and bacteria. This is due to the even distribution of pigments in the archaeal cells, which provides them with more surface area per volume and allows them to effectively shade the more concentrated pigments in the *Dunaliella* cells (Oren *et al.*, 1992).

Another example where nonphotosynthetic pigments dominate the spectral reflectance comes from extremophiles (including chemotrophic thermophiles) at the edges of hot springs, where temperatures can exceed the 73°C temperature limit for photosynthesis (Meeks and Castenholz, 1971). Thermophiles such as *Thermus aquaticus* (Brock and Freeze, 1969), whose pigmentation may be an adaptation to oxidative stress, generate the inner ring of yellow seen in the Grand Prismatic Spring in Yellowstone National Park, USA (Fig. 1d). Pigmented thermophilic chemotrophs, together with carotenoid-bearing cyanobacteria, form a visible color gradient across the spring (Dartnell, 2011). "Watermelon snow" is a designation given to the visible red or pink coloration of snow caused by the green algae *Chlamydomonas nivalis* (Painter *et al.*, 2001; Williams *et al.*, 2003). This organism thrives in high altitudes and polar regions during the summer with the aid of the red carotenoid pigment astaxanthin, which helps protect it from UV radiation and warms the cell by absorbing more incident solar radiation than the surrounding snow.

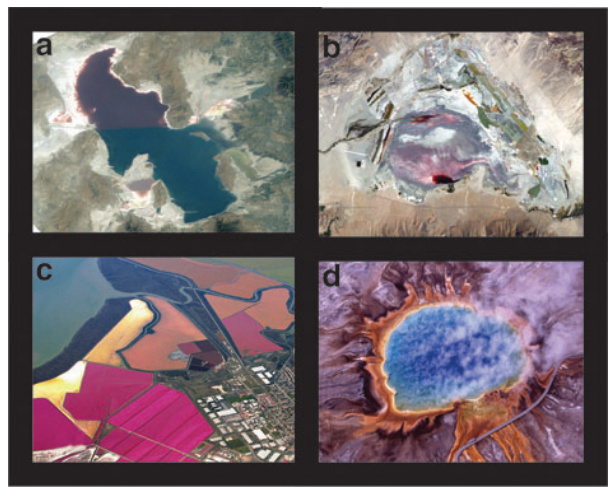

**FIG. 1.** Macroscopic surfaces where carotenoid-type pigments dominate the spectral reflectance. (**a**) The Great Salt Lake, Utah, USA, seen from the International Space Station (credit: NASA). The Great Salt Lake is approximately 120 km long, 45 km wide, and 4.9 m deep on average. (**b**) Owens Lake in California, USA (credit: NASA). Owens Lake is 28 km long, 16 km wide, and 0.9 m deep on average. (**c**) Salt ponds in San Francisco, California, USA (credit: dro!d, Atlanta, GA). (**d**) The Grand Prismatic Spring in Yellowstone National Park, Wyoming, USA (credit: National Park Service). The spring is 90 m long by 80 m wide and approximately 50 m deep.

Many types of vegetation alter their spectral appearance (color) seasonally, illustrating another case where nonphotosynthetic pigments can dominate the visible reflectance spectrum. The red and orange autumn coloration of leaves is due to carotenoid pigments that are unmasked as chlorophyll degrades, while red pigmentation is due to anthocyanin, a pH-dependent pigment that is produced *de novo* in autumn foliage, perhaps to provide photoprotection (Archetti *et al.*, 2009).

Nonphotosynthetic pigments can serve a variety of functions, many of which help the organism adapt to stressors in the environment. These compounds can be used for photoprotection (Proteau *et al.*, 1993; Williams *et al.*, 2003; Solovchenko and Merzlyak, 2008; Archetti *et al.*, 2009) and as quenching agents for protection against free radicals (Saito *et al.*, 1997). Desiccation- and ionizing-radiation-resistant organisms such as *Deinococcus radiodurans* and *Rubrobacter radiotolerans* contain carotenoid pigments that are thought to function primarily as antioxidants (Saito *et al.*, 1994; Cox and Battista, 2005; Tian *et al.*, 2008). Some pigments serve as biocontrol mechanisms to slow growth as resources are exhausted (Venil and Lakshmanaperumalsamy, 2009) or to facilitate interactions between bacterial cells in colonies or aggregates through a phenomenon known as "quorum sensing" (McClean *et al.*, 1997; Williams *et al.*, 2007). Such pigments are often associated with pathogenic bacteria, as invading a host exposes these organisms to extremes in terms of temperature and in many cases an immune system attack (Liu and Nizet, 2009). "Siderophore" pigments are used as $Fe^{3+}$ bonding agents in iron-limited conditions (Meyer, 2000). Plants and animals use pigments in signaling to other organisms (Chittka and Raine, 2006), as is the case for pigments used in flowers for attracting pollinators.

In summary, the spectral properties of many biological pigments are decoupled from the light environment to varying degrees. While some nonphotosynthetic pigments are adapted to be sensitive to small portions of the electromagnetic spectrum, such as UV-screening pigments, others are hosted by organisms in zero light conditions (Kimura *et al.*, 2003). Organisms possessing nonphotosynthetic pigments are not necessarily closely related genetically to photosynthetic species, and the phylogenetic diversity of pigment-producing species in general is much broader than that of photosynthetic species (Klassen, 2010). Table 1 shows a list of functions that pigment molecules perform other than light harvesting and provides specific examples of pigments that carry out these functions and the organisms in which they are found. We note that some carotenoids can have a light-harvesting function in anoxygenic phototrophs and cyanobacteria but predominantly function as protectants against photooxidative stress (Cogdell *et al.*, 2000; Glaeser and Klug, 2005; Ziegelhoffer and Donohue, 2009). For the purposes of this paper, we consider these carotenoids as photoprotectants and antioxidants and thus as nonphotosynthetic pigments.

To expand the study of remotely detectable surface biosignatures beyond those associated with photosynthesis, we explore the nature, diversity, and detectability of nonphotosynthetic pigments. Our study is broken into four components. First, we measure the reflectance spectra of a diverse collection of pigmented organisms on Earth. These organisms may be considered as analogues to the possible biologically pigmented surfaces that may be present on other planets. Second, we generate synthetic spectra with a 1-D radiative transfer model that simulates ideal cases where single surface types dominate planetary surfaces and are viewed through a planetary atmosphere. We compare scenarios with nonphotosynthetic pigment surfaces to those with other dominant surface types. We include in these simulations both the spectral reflectances measured in the first, experimental portion of the study and spectral reflectances of biotic and abiotic surfaces from literature. Third, we create a more comprehensive 3-D spectral model of an Earth-analog planet where the oceans are dominated by pigmented halophiles at the same densities found in high-salinity ponds on Earth and include the effects of an atmosphere, clouds, and a heterogeneous surface. Finally, we examine the utility of

Table 1. Functions of Pigments

| Function | Pigment type(s) | Example pigments | Example organism(s) | References |
|---|---|---|---|---|
| Photosynthesis | Chlorophylls, bacteriochlorophylls | Chls a, b; Bchls a, c, g | Cyanobacteria, anoxygenic phototrophs | Hohmann-Marriott and Blankenship, 2012; Clayton, 1966 |
| Other light capture, other phototrophy | Some rhodopsins, some carotenoids | Xanthorhodopsin, bacteriorhodopsin | *Salinbacter ruber, Halobacterium salinarum* | Boichenko *et al.*, 2006; Oren, 2013; Grote and O'Malley, 2011 |
| Sunscreen | cyclized β-ketoacid | Scytonemin | Cyanobacteria | Proteau *et al.*, 1993 |
| Antioxidant | Carotenoids | Bacterioruberin, deinoxanthin | *Halobacterium salinarum, Deinococcus radiodurans* | Lemee *et al.*, 1997; Saito *et al.*, 1997; Shahmohammadi *et al.*, 1998 |
| Protection against temperature extremes | Tyrosine derivative | Melanin | *Cryptococcus neoformans* | Dadachova *et al.*, 2007; Liu and Nizet, 2009 |
| Acquisition of nutrients such as iron | Siderophore | Pyoverdine | *Pseudomonas putida* | Meyer, 2000 |
| Regulation of growth (cytotoxicity) | Prodiginine | Prodigiosin | *Serratia marcescens* | Bennett and Bentley, 2000; Hejazi and Falkiner, 1997 |
| Protection against competition or grazing (antimicrobial) | Indole derivative | Violacein | *Janthinobacterium lividum* | Kimmel and Maier, 1969; Schloss *et al.*, 2010; Durán *et al.*, 2007 |
| Signaling other organisms | Carotenoids, anthocyanins, betalains | Cryptoxanthin | *Narcissus pseudonarcissus* | Tanaka *et al.*, 2008; Chittka and Raine, 2006; Valadon and Mummery, 1968 |
| Bioluminescence | Luciferin | Dinoflagellate luciferase | Dinoflagellates | Haddock *et al.*, 2010 |

The diverse functions of biological pigments.

broadband colors in identifying and characterizing a diversity of biological and abiotic surface types when full consideration of atmospheric effects is made.

## 2. Methods and Models

Below we will describe the methods and the models for both the experimental and spectral modeling components of this study.

### 2.1. Choice of cultured organisms

We cultured and measured the spectral characteristics of a variety of pigmented bacteria (and one archaeon) that represent various colorations, metabolisms, and phylogenies. For practical reasons, we preferentially selected fast-growing, easily culturable strains. Table 2 lists the species, phylum, color, primary pigment, metabolism, and environment of the selected organisms. Below we briefly describe each organism.

*Brevibacterium aurantiacum*—an orange Gram-positive actinobacterium that has been isolated from food products, rice plots, and industrial waste. The production of orange carotenoid pigmentation is observed to be induced by blue (~400 nm) light (Gavrish *et al.*, 2004; Takano *et al.*, 2006).

*Chlorobium tepidum*—a green sulfur Gram-negative bacterium that grows photoautotrophically or photoheterotrophically under anaerobic conditions. *C. tepidum* was isolated from an acidic (pH 4.5–6.0) hot spring containing high concentrations of sulfide (Wahlund *et al.*, 1991). *Chlorobium* species are often found in anoxic zones of eutrophic lakes

(Prescott *et al.*, 2005). Molecular fossils (biomarkers) indicate *Chlorobium* blooms were present in high numbers during the end-Permian mass extinction event (Cao *et al.*, 2009).

*Deinococcus radiodurans*—a well-known polyextremophile that stains Gram-positive (although its cell wall has Gram-negative characteristics), which can survive exposure to large amounts of ionizing radiation and desiccating conditions (Cox and Battista, 2005). The primary pigment produced by *D. radiodurans* is the carotenoid deinoxanthin, which gives colonies a red-orange coloration.

*Halobacterium salinarum*—a Gram-negative halophilic archaeon that contains the red carotenoid pigment, bacterioruberin, and achieves photoheterotrophy with bacteriorhodopsin. Bacterioruberin has been shown to assist *H. salinarum* in resisting damage to DNA by ionizing radiation and UV light (Shahmohammadi *et al.*, 1998). *Halobacterium* species are of interest to astrobiologists because they are polyextremophiles and occupy a niche populated by few other organisms (DasSarma, 2006). *H. salinarum* was chosen in his work to represent the pigmented halophiles that give hypersaline lakes and ponds their pink and red color.

*Janthinobacterium lividum*—a Gram-negative aerobic proteobacterium commonly found in soils, freshwater, and tundra. *J. lividum* produces the dark violet pigment violacein and can aggregate in dense colonies or biofilms (Kimmel and Maier, 1969; Pantanella *et al.*, 2007; Schloss *et al.*, 2010). The pigment violacein is involved in "quorum sensing," which can facilitate interactions between cells in a colony (McClean *et al.*, 1997; González and Keshavan, 2006; Williams *et al.*, 2007).

TABLE 2. Cultured Species: Classifications, Pigments, Metabolisms, and Environments

| Species | Phylum | Visual color | Pigment(s) | Metabolism(s) | Environment(s) | References |
|---|---|---|---|---|---|---|
| Brevibacterium aurantiacum | Actinobacteria | Orange | 3,3-Dihydroxy-isorenieratene or canthaxanthin | Aerobic heterotroph | Food products, rice plots, industrial waste | Gavrish et al., 2004 |
| Chlorobium tepidum | Chlorobi | Green | Bchls a, c; chlorobactene | Anaerobic photolithotroph | Anoxic zones of eutrophic lakes | Wahlund et al., 1991 |
| Deinococcus radiodurans | Deinococcus-Thermus | Red-orange | Deinoxanthin | Aerobic heterotroph | Deserts, desiccating environments | Cox and Battista, 2005; Tian et al., 2008 |
| Halobacterium salinarum | Euryarchaeota | Pink-red | Bacterioruberin; bacteriorhodopsin | Aerobic photoheterotroph | Hypersaline lakes, salterns | DasSarma, 2006; Oren, 2009; Shahmohammadi et al., 1998 |
| Janthinobacterium lividum | Proteobacteria | Dark-violet | Violacein | Aerobic heterotroph | Soils, tundra | Kimmel and Maier, 1969; Schloss et al., 2010 |
| Micrococcus luteus | Actinobacteria | Yellow | Canthaxanthin (a $\beta$-carotene); $\alpha$-carotenes | Aerobic heterotroph | Soil, air, water | Ungers and Cooney, 1968; Greenblatt et al., 2004 |
| Phaeobacter inhibens | Proteobacteria | Brown | Unknown | Aerobic heterotroph | Sea water | Porsby et al., 2008; Martens et al., 2006; Dogs et al., 2013 |
| Rhodobacter sphaeroides | Proteobacteria | Brown-green | Bchls a, b; spheroidenone; neurosporene | Anaerobic photoautotroph, aerobic chemoheterotroph | Deep lakes, stagnant waters | Pfennig and Truper, 1971; Mackenzie et al., 2007; Wiggli et al., 1999; Glaeser and Klug, 2005 |
| Rhodopseudomonas palustris | Proteobacteria | Red-purple | Bchl a, neurosporene | Various—photoautotrophic, photoheterotrophic, chemoheterotrophic, chemoautotrophic | Pond water, coastal sediments | Larimer et al., 2004; Bosak et al., 2007 |
| Rubrobacter radiotolerans | Actinobacteria | Red | Bacterioruberin, monoanydrobacterioruberin | Aerobic heterotroph | Desert soil | Saito et al., 1994; Schabereiter-Gurtner et al., 2001; Chen et al., 2004 |
| Serratia marcescens | Proteobacteria | Red | Prodigiosin | Aerobic heterotroph | Soils, household surfaces, animal tissue | Héjazi and Falkiner, 1997; Bennett and Bentley, 2000; Schuerger et al., 2013; Haddix et al., 2008 |

Data for cultured organisms (Section 2.1).

Table 3. Culture Information

| Species | DSMZ strain ID # | DSMZ media # | Culture type[a] |
|---|---|---|---|
| *Brevibacterium aurantiacum* | 20426 | 92 | Aerobic; plated |
| *Chlorobium tepidum* | 12025 | 29 | Anaerobic in light; liquid |
| *Deinococcus radiodurans* | 20539 | 53 | Aerobic; plated |
| *Halobacterium salinarum* | 22414 | 97 | Aerobic (37°C); plated |
| *Janthinobacterium lividum* | 1522 | 1, 1a | Aerobic; plated |
| *Micrococcus luteus* | 20030 | 1 | Aerobic; liquid |
| *Phaeobacter inhibens* | 16374 | 514 | Aerobic; liquid |
| *Rhodobacter sphaeroides* | 158 | 27 | Anaerobic in light; liquid |
| *Rhodopseudomonas palustris* | 8283 | 651 | Anaerobic in light; liquid |
| *Rubrobacter radiotolerans* | 5868 | 535 | Aerobic; liquid |
| *Serratia marcescens* | 30121 | 1 | Aerobic; liquid |

Data for cultured strains. Aerobic cultures were grown in the presence of oxygen, whereas anaerobic cultures were grown in airtight bottles evacuated of oxygen. Further detailed information regarding strain type and culture media can be found online at https://www.dsmz.de.

[a]All cultures were grown at 25°C unless noted otherwise.

*Micrococcus luteus*—a Gram-positive, yellow-pigmented aerobic actinobacterium. The yellow pigmentation is due to the carotene derivative canthaxanthin (Ungers and Cooney, 1968).

*Phaeobacter inhibens*—a Gram-negative brown-pigmented proteobacterium isolated from seawater. *P. inhibens* produces an uncharacterized brown pigment and displays antimicrobial properties (Porsby *et al.*, 2008; Dogs *et al.*, 2013).

*Rhodobacter sphaeroides*—a Gram-negative purple non-sulfur anoxygenic photosynthetic proteobacterium. *R. sphaeroides* has a diverse array of metabolic pathways, growing by aerobic or anaerobic respiration, photosynthesis or fermentation (Mackenzie *et al.*, 2007). Carotenoids produced by *R. sphaeroides* have been shown to reduce photooxidative stress against singlet oxygen (Glaeser and Klug, 2005) and are the cause of the orange-brown coloration the organism exhibits in anaerobic cultures. The *R. sphaeroides* strain in this study was grown photoheterotrophically in anaerobic conditions.

*Rhodopseudomonas palustris*—a Gram-negative purple anoxygenic photosynthetic proteobacterium. Like *R. sphaeroides*, *R. palustris* is metabolically versatile, able to grow photoautotrophically, photoheterotrophically, chemoautotrophically, or chemoheterotrophically (Larimer *et al.*, 2004). This diverse array of metabolisms allows the organisms to survive in a variety of environments. *R. palustris* is especially interesting as it has been shown to precipitate calcite in solutions rich in calcium carbonate, serving as a model to illuminate how ancient anoxygenic photosynthetic stromatolites may have been constructed (Bosak *et al.*, 2007). The *R. palustris* strain in this study was grown photoheterotrophically in anaerobic conditions.

*Rubrobacter radiotolerans*—a Gram-positive red-pigmented actinobacterium that is isolated from desert soil. *R. radiotolerans* is highly radioresistant and can survive radiation doses higher than other radioresistant bacteria including *Deinococcus radiodurans* (Saito *et al.*, 1994). *R. radiotolerans* contains bacterioruberin and other pigments typically found in halophilic bacteria (Saito *et al.*, 1994).

*Serratia marcescens*—a common Gram-negative environmental proteobacterium that produces the red pigment prodigiosin (Venil and Lakshmanaperumalsamy, 2009). *S. marcescens* is often found growing as a thin pink film on bathroom surfaces and has been known to cause infections (Hejazi and Falkiner, 1997; Haddix *et al.*, 2008). A related organism, *Serratia liquefaciens*, which also produces prodigiosin, has been shown to grow in an anoxic, $CO_2$-enriched 7 mbar (Mars-like) atmosphere at 0°C (Schuerger *et al.*, 2013).

## 2.2. Growth and harvesting of cells

Each organism in Table 2 was acquired from Deutsche Sammlung von Mikroorganismen und Zellkulturen (DSMZ[1]) in the form of either a live culture or a freeze-dried pellet in a vacuum-sealed glass ampoule. Standard culturing procedures were followed after assembling the recommended growth medium for each organism (see Table 3). All species were grown in oxic conditions with the exception of the anaerobic anoxygenic photosynthesizers *Chlorobium tepidum*, *Rhodobacter sphaeroides*, and *Rhodopseudomonas palustris*, which were grown anaerobically in light. Cells were harvested for reflectance spectra measurements in one of two ways:

(1) Liquid cultures were transferred to 50 mL Falcon tubes and centrifuged for 10 min at 2700*g* to pelletize the intact cells. The supernatant was removed and replaced with deionized water. The tube was agitated to rinse the cells of the growth media and then centrifuged again for 10 min at 2700*g* to re-isolate the cells as a pellet. A sterilized small metal spatula was used to transfer the cleaned cell pellet to black filter paper in a Petri dish (black paper was chosen to minimize the scattering of light during subsequent reflectance spectra measurements).

(2) Plate-spread cultures were allowed to proliferate until a substantial percentage of the plate was covered with growth. A sterilized metal spatula was then used to gently scrape the surface of the growth media until a thick paste of cells was gathered. This was transferred to filter paper in a Petri dish.

In both (1) and (2), the sample was allowed to dry for 1 h to remove any film of water on the surface of the sample that could cause specular reflection. The resulting cell mass was at least $0.5 \times 0.5$ cm and opaque. See Fig. 2 for an example. Growth media and dead cell matter were removed because these factors varied from strain to strain and would complicate the direct comparison of one organism to another. Their removal ensured that the spectral features observed in the subsequent reflectance spectra measurements originated from the pigmented cells.

---

[1]https://www.dsmz.de

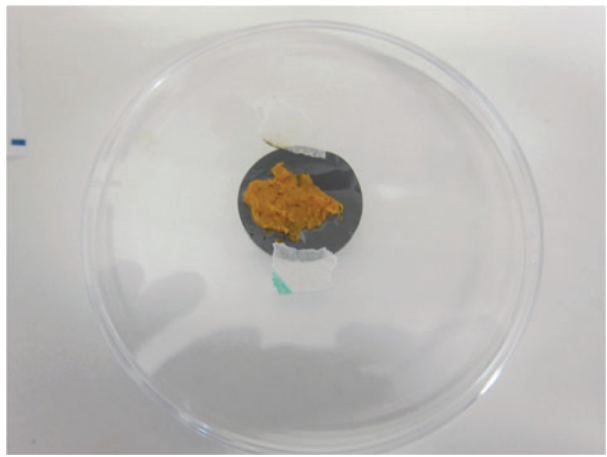

**FIG. 2.** A sample of *Brevibacterium aurantiacum* before measurement with the Ocean Optics spectrometer.

### 2.3. Spectral reflectance measurements

Spectral reflectance measurements were made of a monoculture of each species to determine the wavelength-dependent features of the organism's reflectance spectrum without environmental effects. Specifically, we aimed to quantify the location and strength of major spectral features. We assumed the samples were sufficiently thick to be diffuse reflectors due to multiple scattering within the cellular layers (see, *e.g.*, Broschat *et al.*, 2014). The light environments are similarly diffuse above microbial mats mixed within a sediment matrix (Jorgensen and Des Marais, 1988; Kühl *et al.*, 1997; Decho *et al.*, 2003).

Reflectance spectra measurements were made with an Ocean Optics USB2000+ UV/vis spectrometer (Ocean Optics grating #3) with 2048 pixel channels, a dispersion of 0.32 nm per pixel, and an optical resolution of 4.12 nm. The spectrometer was attached to a reflectance probe (Ocean Optics R400-7) via a 400 $\mu$m diameter SMA 905 fiber optic cable. The reflectance probe contained one read fiber and six illuminating fibers arranged in a circular pattern around the read fiber. An Ocean Optics HL-2000 tungsten-halogen light source was joined to the fiber optic cable. A similar experimental configuration was described by Decho *et al.* (2003).

The reflectance probe was secured by an adjustable clamp attached to a vertical ring stand. Samples were placed on a custom holder on a level surface below the reflectance probe. The probe's position was maintained 30 mm from the surface of the sample with a 0° (perpendicular) orientation. The entire sampling apparatus was contained within a custom-built dark box to exclude ambient light. This box was coated in matte black paper to minimize scattering of light from the light source.

The wavelength-dependent radiance reflectance ($R(\lambda)$) was calculated according to the following equation:

$$R(\lambda) = \frac{C(\lambda)_{\text{sample}} - C(\lambda)_{\text{dark}}}{C(\lambda)_{\text{standard}} - C(\lambda)_{\text{dark}}} \times \frac{t_{\text{standard}}}{t_{\text{sample}}} \times k(\lambda)_{\text{WS}-1} \quad (1)$$

where $C(\lambda)_{\text{sample}}$ is the counts from the sample measurement, $C(\lambda)_{\text{dark}}$ is the dark count with no illumination, $C(\lambda)_{\text{standard}}$ is the counts measured from the standard, $t_{\text{sample}}$ is the integration time of the sample, $t_{\text{standard}}$ is the integration time of the standard, and $k(\lambda)_{\text{WS}-1}$ is the wavelength-dependent calibration factor for the WS-1 diffuse reflectance standard. The factor $k(\lambda)_{\text{WS}-1}$ varies between 0.992 and 0.993 for the wavelength range measured in this work (tables with the wavelength-dependent calibration factors are available on the Ocean Optics Web site[2]).

Calibrations for reflectance measurements were made as follows. The light source was turned on several minutes before calibration measurements were made. The Spectralon WS-1 diffuse reflectance standard was placed in the sample position 30 mm below the reflectance probe. A series of 100–500 calibration spectra with integration times of 1 ms were collected from the illuminated reflectance standard. These calibration spectra were averaged and stored to computer memory. To measure the dark current, the light was blocked from entering the fiber with a switch-operated internal shutter, and a complementary series of calibration spectra were taken and stored to memory. New dark and reflectance calibration spectra were taken before each new sample measurement set.

Three to five spectral sets were recorded for each organism sample. Each set was a series of 100–500 spectral scans encompassing the 0.4–0.85 $\mu$m wavelength range. Each scan had an integration time of 1 ms, chosen to achieve maximum signal-to-noise without saturating the detector. The series of spectral scans were then averaged to produce a composite spectrum. This procedure was repeated 3–5 times at different locations on the sample surface to produce a spectral set. The location of the sample under the fiber was adjusted for each composite spectrum in the set, such that the sensor viewed a smooth area of the sample to minimize the effect of surface roughness on the reflectance of the isolated cells. The final spectrum for each organism was computed by taking the median value at each wavelength point of the averaged composite spectra in the spectral set. This setup yields spectra with high signal-to-noise for the 0.4–0.85 $\mu$m spectral range. The standard deviation for any wavelength in the composite spectra was less than 5%. Although the fiber-spectrometer combination had spectral sensitivity from 0.35 to 0.85 $\mu$m, shorter wavelengths were inaccessible because the tungsten-halogen fiber optic light source does not produce significant UV light.

### 2.4. Spectral modeling of simulated planets

In the realistic case of a disk-averaged planetary observation, a pigmented surface will be viewed through an atmosphere; therefore, a remote observation of a planetary spectrum will include atmospheric absorption and scattering effects by gaseous molecules and clouds. To explore the effects of the atmosphere on the detectability of reflectance from surface pigments, we use a radiative transfer model to calculate top-of-atmosphere spectra for both abiotic and biotic surfaces observed through an Earth-like atmosphere. We use the abiotic surfaces to compare spectra where biosignatures are present to those cases where they are not present.

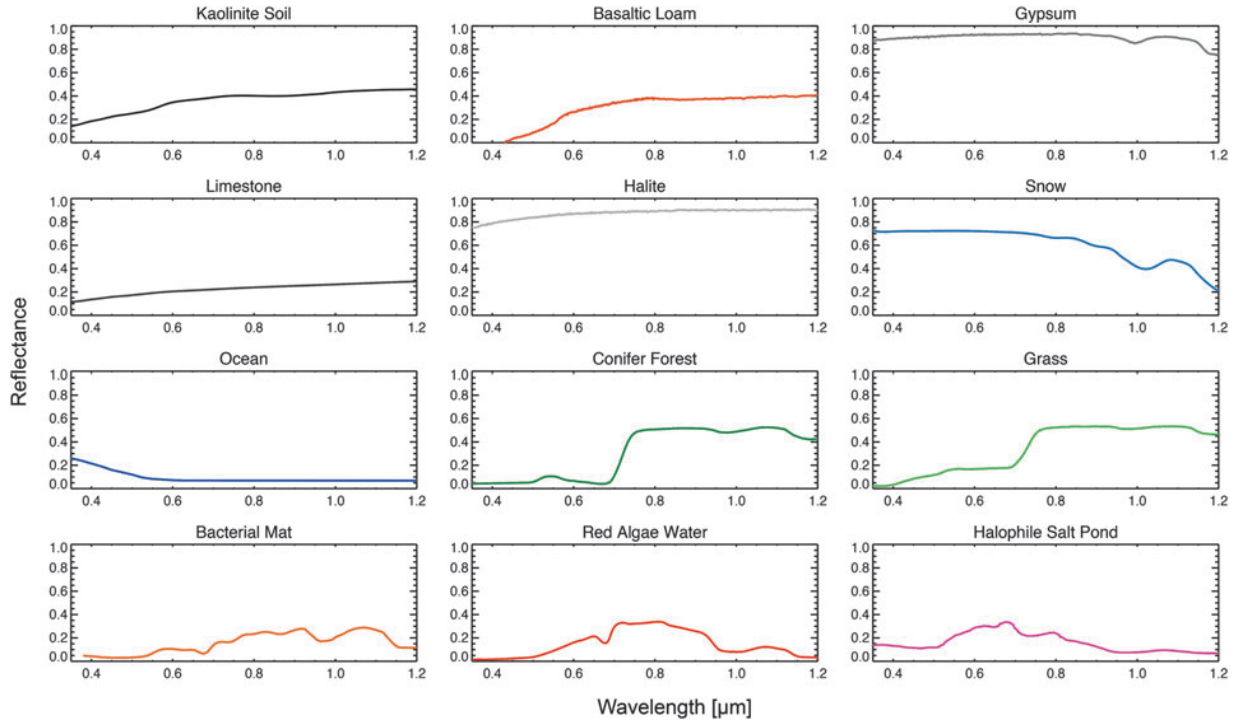

**FIG. 3.** Spectral reflectances of abiotic and biotic surfaces from the USGS (Clark *et al.*, 2007); and ASTER (Baldridge *et al.*, 2009) spectral libraries from 0.35 to 1.2 μm. The halophile salt pond's spectral reflectance is from work by Dalton *et al.* (2009). Note the water absorption features in the bacterial mat, red algae water, and halophile salt pond cases.

We use the reflectance spectra measured in Section 2.3 (and presented in Section 3.1) as input surface reflectances for one series of simulated spectra. Another series of spectra uses spectral reflectances obtained from the literature and encompasses a wavelength range that is larger than our experimental spectra. Figure 3 shows the wavelength-dependent reflectances of kaolinite soil, basaltic loam, gypsum, limestone, halite, snow, ocean water, a conifer forest, grass, a bacterial mat, red algae–coated water, and a halophile-dominated saltern evaporation pond. The reflectance spectrum for the halophile-dominated pond (see Fig. 1c) was taken from the work of Dalton *et al.* (2009), which provides reflectance spectra of the San Francisco salt ponds measured *in situ* and therefore includes the spectral effects of both the pigmented organisms contained within the pond and the overlying layers of water. The other surface reflectances are from the USGS and ASTER spectral libraries (Clark *et al.*, 2007; Baldridge *et al.*, 2009). The conifer forest and the halophile-dominated pond provide a comparison between surfaces dominated by photosynthetic and nonphotosynthetic organisms that are known to exist on Earth. The organisms that constitute these biological surfaces contain pigments that cause reflectance peaks at different wavelengths and thus may be distinguishable from each other.

We use our 1-D model to generate a spectrum of a planet dominated by each of the surfaces listed above under an Earth-like atmosphere in a clear-sky (no cloud) scenario. We then use a 3-D spectral Earth model with clouds, which simulates the disk-averaged spectrum of the heterogeneous and time-dependent Earth with high fidelity (Tinetti *et al.*, 2006a; Robinson *et al.*, 2011, 2014), to explore a more realistic scenario comparing the modern (simulated) Earth with a plausible Earth-like planet containing a salty ocean dominated by nonphotosynthetic pigmented organisms.

#### 2.4.1. One-dimensional clear-sky models.

In this case, we consider a planetary surface dominated by a single surface type possessing either the experimentally measured spectral reflectances or previously known spectral reflectances from literature. We model these surfaces under an annually averaged midlatitude Earth-like atmosphere and an incident stellar spectrum identical to the Sun's. We assume the surfaces are diffuse reflectors, which has been demonstrated to be an adequate assumption when comparing synthetic spectra to validation data for Earth at gibbous phases (Williams and Gaidos, 2008; Robinson *et al.*, 2010; Cowan and Strait, 2013). We assume an Earth-like atmosphere including average temperature-pressure and gas mixing ratio profiles (Earth-like atmospheres are assumed because each biological and nonbiological surface exists on modern Earth). The spectrally active (absorptive) gases included in the model are $H_2O$, $CO_2$, $O_3$, $O_2$, $N_2O$, CO, and $CH_4$. The 1-D radiative transfer model used is the Spectral Mapping Atmospheric Radiative Transfer Model (SMART) developed by D. Crisp (Meadows and Crisp, 1996; Crisp, 1997). The HITRAN 2008 (Rothman *et al.*, 2009) line lists were used to calculate the absorption cross-sections for each spectrally active gas. The resolution of the spectral grid in the model was $1\,cm^{-1}$, which corresponds to a wavelength resolution of $\Delta\lambda = 2.5 \times 10^{-5}\,\mu m$ at $\lambda = 0.5\,\mu m$. We used a single solar zenith angle of 60°, which approximates the illumination observed in a planetary disk average (Segura *et al.*, 2005), and consider cases with no cloud cover. This mimics a clear sounding through the atmosphere and is only achievable for a partially cloud-covered planet when the observation has adequate spatial resolution. Realistically, some level of cloud cover is expected, and while obtaining longitudinally resolved observations of terrestrial exoplanets may be possible with adequately time-resolved photometry from future space-based telescopes (Cowan *et al.*, 2009, 2011; Cowan and Strait, 2013), the spatial resolution obtainable will span a significant fraction of the planet and so will likely still include clouds. While this will affect absolute detectability of surface reflectance biosignatures, it will not introduce spurious wavelength-dependent features in the spectrum because clouds are approximately gray (*i.e.*, they have largely wavelength-independent reflectivity) in the visible regime. The spectrum of a partially cloud-covered planet will be a linear combination of spectra from clear soundings and spectra from fully and partially cloud-covered soundings. The effect on the planetary spectrum of spatially inhomogeneous clouds is more fully captured in the 3-D model described below.

### 2.4.2. Three-dimensional spectral Earth model.

To better assess the detectability of the most promising nonphotosynthetic pigmented organism in a plausible context, and to compare with spatially resolved models of inhabited Earth-like planets in the literature (*e.g.*, Tinetti *et al.*, 2006b; Robinson *et al.*, 2011; Sanromá *et al.*, 2013, 2014), we produced more comprehensive whole-planet spectra that contain the confounding effects of spatially resolved clouds and different surface types. To do this we use the Virtual Planetary Laboratory's 3-D spectrally resolved Earth model (described by Robinson *et al.*, 2011). Briefly, this model incorporates data from Earth-observing satellites, including spatially dependent snow and cloud cover, surface temperature, and gas mixing ratio profiles, as input to a spatially and altitudinally resolved radiative transfer model of a distant Earth-like planet. These properties were interpolated from high-spatial-resolution onto lower-resolution maps by using the Hierarchical Equal Area and isoLatitude Pixelization (HEALPix) method (Gorski *et al.*, 2004). For the model runs presented here, the atmospheric parameters varied spatially over 48 atmospheric pixels with 40 vertical layers, and the surface was approximated as 192 pixels. The surface reflectance of each surface pixel was calculated as a linear combination of five surface types (ocean, soil, snow, forest, and grassland). The modern Earth continental arrangement is used for the land and ocean spatial distribution. We assume the observer is viewing the disk-averaged planet from an approximately sub-equatorial latitude (1.6°N) and a Sun-planet-observer phase angle of 57.7° over a 24 h period from March 18–19. This date was chosen because the Virtual Planetary Laboratory's Earth model has already been validated on this date with observed near-infrared and broadband visible spectra taken during the EPOXI mission by the Deep Impact spacecraft (Livengood *et al.*, 2011;

Robinson *et al.*, 2011). The disk-integrated spectrum of the planet produced by the model was divided by a solar spectrum corrected for phase (*i.e.*, partial illumination of the planet) with a Lambertian phase function, to produce the whole-disk planetary spectral albedo (hereafter referred to as the ''albedo''). It has been shown that Earth's phase function only deviates significantly from a Lambertian function near crescent phase (phase angles near zero) where glint (specular reflection from the ocean) and cloud forward scattering become important (Williams and Gaidos, 2008; Robinson *et al.*, 2010).

To explore the potential detectability of a nonphotosynthetic pigment with as comprehensive a model as possible, we used the 3-D spectral Earth model described above to compare the spectrum of a realistic Earth with a similar model case in which the spectral reflectance of the oceans has been replaced by that of the halophile-dominated saltern pond. This environment was chosen for further modeling because the halophile spectrum produced a strong spectral signature that was comparable in strength to that of photosynthetic plants (*e.g.*, the conifer forest), and so had a higher probability of being detectable in the disk-average. This simulation approximates a scenario in which an Earth-like planet contains a very salty, shallow ocean populated by pigmented halophilic organisms at the same density found in the San Francisco saltern ponds with the highest salinity (Dalton *et al.*, 2009), but with the same cloud and snow/ice cover of modern Earth. This ''halophile Earth'' case presents a best-case end-member scenario for the surface coverage of pigmented halophilic organisms on an Earth-like planet.

## 2.5. Broadband colors

Broadband colors were calculated to investigate their usefulness in identifying exoplanet surface types via astronomical observations and to search for possible patterns when comparing the colors produced by biotic or abiotic surfaces. Broadband color in this context is the difference in brightness between two bands. In astronomical observations, this is reported as a logarithm. To compare two bands $X$ and $Y$, we used the equation

$$C_{XY} = X - Y = -2.5 * \log_{10}\left(\frac{r_X}{r_Y}\right) \qquad (2)$$

where $r_X$ is the reflectivity in band $X$ and $r_Y$ is the reflectivity in band $Y$. We define reflectivity here as the irradiance from the planet in a given band divided by the irradiance of the incident solar spectrum. Typically in astronomical observations color is a difference in observed flux between bands, which will vary as a function of the host star's spectrum when the flux from the target object is composed of reflected light from the star. Here, we use reflectivity only in calculating broadband colors and assume that the host star's spectrum, which will be known, has been divided out of the observed planet flux. We first calculate broadband colors of the experimentally measured reflectance spectra, and the biotic and abiotic surfaces from spectral libraries, neglecting the effects from atmospheric absorption and scattering. We then calculate the colors using our simulated spectra (Section 2.4) that account for the effects of viewing a surface through a planetary atmosphere. We define three bands for our illustration: B* = Blue = 0.4–0.5 μm, V* = Visible = 0.5–0.7 μm, and I* = Infrared = 0.7–0.85 μm. These bands are modeled after the Johnson-Cousins BVI bands but differ in that they have perfect transmission through the entire idealized filter band-pass for comparison with other studies of broadband colors of planetary surfaces (*e.g.*, Hegde and Kaltenegger, 2013). The infrared band additionally has a smaller width and terminates at a shorter wavelength (0.85 μm vs. ~0.9 μm) due to limitations on our measurements from the Ocean Optics spectrometer.

## 3. Results

### 3.1. Reflectance spectra measurements

The reflectance spectrum for each organism we measured in the laboratory is shown in Fig. 4. The wavelength-dependent spectral reflectance of a conifer forest, obtained from the ASTER spectral library (Baldridge *et al.*, 2009), is provided for comparison. The strength of spectral break features such as the VRE can be quantified by measuring the reflectance change between the maximum reflectance and the nearest local minimum in reflectance, given here as

$$\Delta R = R_{\lambda_{max}} - R_{\lambda_{min}} \tag{3}$$

where $R_{\lambda_{max}}$ is the maximum reflectance and $R_{\lambda_{min}}$ is the reflectance at the nearest local minimum. Another way to quantify the strength of the spectral feature is to calculate the change in reflectance factor ($f_{\Delta R}$) over a wavelength interval:

$$f_{\Delta R} = \frac{R_{\lambda_{max}} - R_{\lambda_{min}}}{R_{\lambda_{min}}} = \frac{\Delta R}{R_{\lambda_{min}}} \tag{4}$$

Table 4 reports the locations of notable spectral features in our measured reflectance spectra and the strength of those features as defined above. The strength of the red edge for a conifer forest, an increase in reflectance from 0.69 to 0.77 μm of $\Delta R \sim 0.5$ and $f_{\Delta R} \sim 12$, is in the upper range of red edge strength. Other vegetation and photosynthetic organisms may have substantially weaker red edges (Kiang *et al.*, 2007a). As shown in the figure and table, there is a diverse range of spectral features in the measured organisms, including edge features that are well within a factor of 2 of the conifer forest red edge strength. Especially notable is *H. salinarum*, which has $\Delta R \sim 0.3$ and $f_{\Delta R} \sim 4$ over a wavelength interval comparable to the conifer forest red edge rise.

We split the spectra into three categories based on the strength and location of major spectral features (also shown in Table 4). The first category includes spectra with a strong spectral break or reflectance increase into the red or near infrared. This category includes the red-edge producing conifers and the bacterioruberin-containing

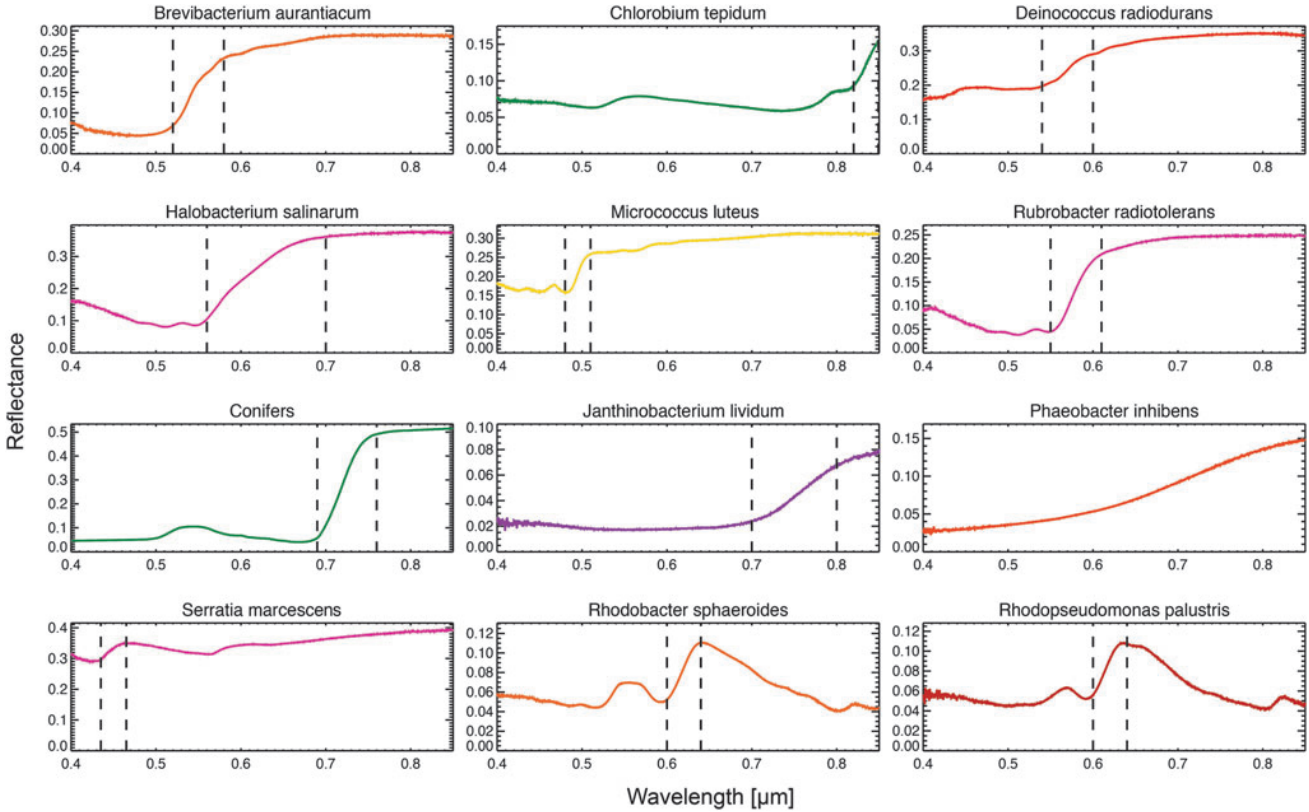

**FIG. 4.** Spectra of all 11 pigmented microorganisms studied in this work organized and grouped into the categories given in Table 4. The seventh panel is the reflectance spectrum of a conifer forest taken from the ASTER spectral library (Baldridge *et al.*, 2009). Spectra were taken from 0.4 to 0.85 μm. The dashed lines bracket the wavelength region containing the strongest spectral feature (listed in Table 4). Note that the vertical axes have different scales in order to show the spectral features of all organisms.

TABLE 4. EMPIRICALLY MEASURED SPECTRAL EDGES AND FEATURES 0.4–0.85 μm

| Organism | "Edges" or spectral breaks (μm) | Absorption features (μm) | Maximum reflectance change (ΔR) | Maximum change in reflectance factor (f$_{ΔR}$) |
|---|---|---|---|---|
| Category 1—Strong Spectral Break and Reflective in the Near IR | | | | |
| B. aurantiacum | 0.52–0.58 | 0.40–0.50 | 0.248 | 5.70 |
| C. tepidum | 0.82–0.86 | 0.52, 0.77 | 0.096 | 1.65 |
| D. radiodurans | 0.54–0.58 | 0.51–0.55 | 0.197 | 1.26 |
| H. salinarum | 0.56–0.70 | 0.48–0.55 | 0.298 | 3.70 |
| M. luteus | 0.49–0.510 | 0.45, 0.485 | 0.158 | 1.01 |
| R. radiotolerans | 0.56–0.60 | 0.51, 0.55 | 0.213 | 5.64 |
| Conifers | 0.69–0.76 | 0.40–0.50, 0.60–0.68 | 0.473 | 11.71 |
| Category 2—Gradual Rise into the Near IR | | | | |
| J. lividum | 0.70–0.80 | 0.40–0.70 | 0.064 | 3.90 |
| P. inhibens | — | — | 0.127 | 5.77 |
| S. marcescens | 0.44–0.46 | 0.41–0.43, 0.57 | 0.107 | 0.37 |
| Category 3—Strong Features in Visible and Absorptive in Near IR | | | | |
| R. sphaeroides | 0.54–0.57, 0.61–0.64 | 0.52, 0.60, 0.805, 0.85 | 0.07 | 1.75 |
| R. palustris | 0.54–0.57, 0.61–0.64 | 0.52, 0.60, 0.805, 0.85 | 0.067 | 1.61 |

Spectral features in measured organisms. "Edges" or spectral breaks (column 2) refer to sharp changes in reflectivity over a small wavelength range. Absorption features can be seen as negative excursions in reflectance.

*Halobacterium salinarum* and *Rubrobacter radiotolerans*. The second category includes spectra that show a shallow or gentle rise into the near infrared, including *Janthinobacterium lividum* and *Phaeobacter inhibens*. The final category includes spectra with strong features in the visible spectrum (<0.7 μm) that are also absorptive in the near infrared, in contrast to Categories 1 and 2. The members of this category of organisms are the anoxygenic photosynthesizers *Rhodobacter sphaeroides* and *Rhodopseudomonas palustris*, which possess bacteriochlorophylls that absorb at near-infrared wavelengths, but also show spectral features in the visible spectrum due to carotenoid pigments.

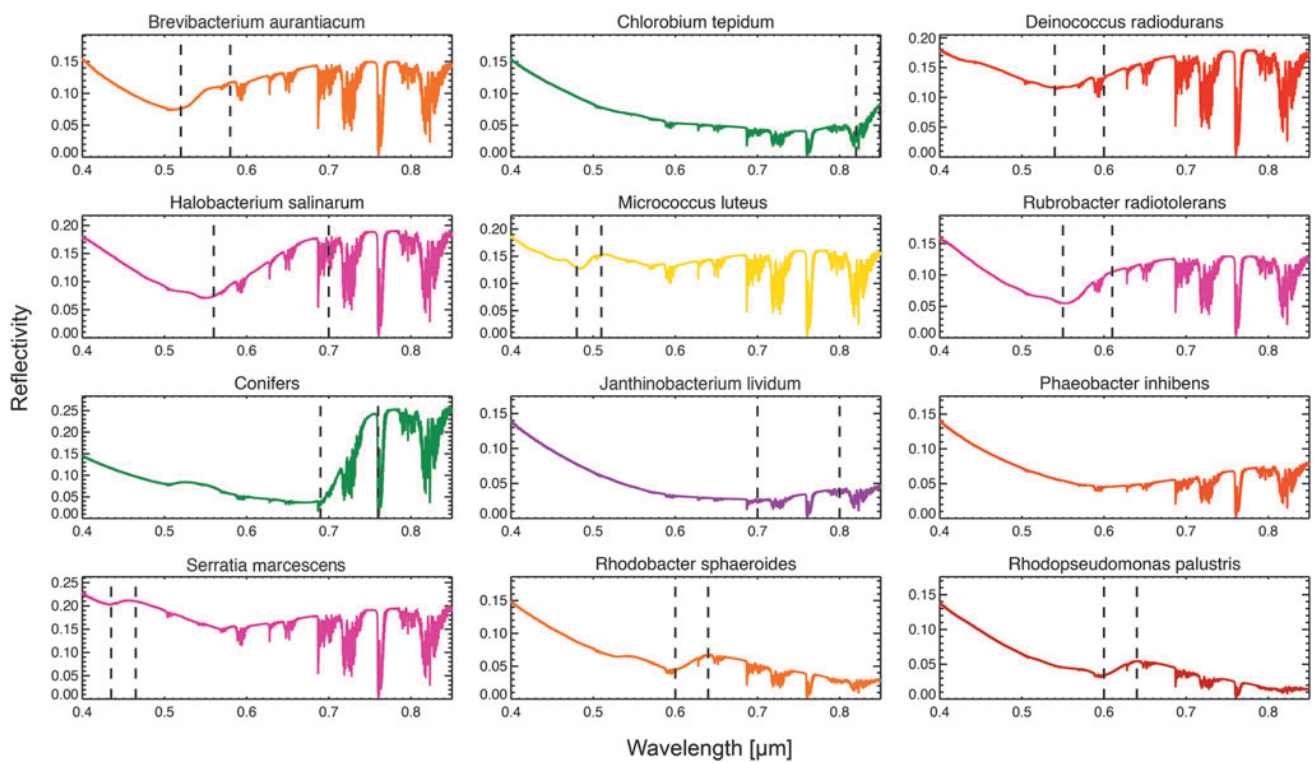

FIG. 5. Resulting top-of-atmosphere spectral reflectivity of planets with surfaces dominated by the pigmented microbes described in Sections 2.3 and 3.1 and shown in Fig. 4. The microbe surfaces are modeled as diffuse reflectors. The average solar zenith angle is assumed to be 60°. The reflectivity is calculated as the modeled top-of-atmosphere spectral irradiance divided by the spectral irradiance of the Sun and not corrected for solar zenith angle. The dashed lines bracket the wavelength region containing the strongest feature seen in the measured reflectance spectrum (listed in Table 4). For the purpose of illustration, the spectra have been smoothed with a moving 5-point statistical mean.

### 3.2. One-dimensional clear-sky models

Figure 5 shows the simulated spectral reflectivity (the top-of-atmosphere irradiance divided by the incident solar irradiance) for sunlight transmitted through an Earth-like atmosphere and reflected from surfaces that have the same diffuse spectral reflectance properties as the experimentally measured microbes. Dashed vertical lines indicate the wavelength region containing the strongest spectral feature in the experimental reflectance spectra. Table 5 lists the maximum fractional reflectivity change in the synthetic spectra over these wavelength regions by using Eq. 4. Because the spectra were generated assuming the average solar zenith angle for a planet observed at quadrature, the reflectivity given in Fig. 5 is not on the same absolute scale as the reflectance in Figs. 3 and 4.

Figure 6 shows the resulting top-of-atmosphere spectral reflectivity for cases of planets with Earth atmospheres and underlying surfaces dominated by kaolinite soil, basaltic loam, gypsum, limestone, halite, snow, ocean water, conifer forest, grass, bacterial mat, red algae–coated water, and a halophile-dominated saltern evaporation pond (surface spectral reflectances shown in Fig. 3). These spectra are presented to show the best-case scenario for detection of that surface type through an Earth-like atmosphere. In Fig. 7, we show for direct comparison the spectra of the conifer forest, halophile salt pond, and ocean case from Fig. 6. The red-edge feature of the forest case is apparent through the cloud-free atmosphere. The spectrum from the halophile salt pond case preserves the $\sim 0.68\,\mu m$ spectral reflectance peak (a 40% increase from a local minimum at $0.57\,\mu m$) and is much more reflective, by a factor of $\sim 3$, than the ocean case from 0.6 to $0.7\,\mu m$.

### 3.3. Three-dimensional models with realistic clouds

We present below the 3-D Earth model synthetic spectra of the realistic Earth and the Earth with halophile-dominated oceans as described in Section 2.4.2. We provide the 24 h diurnally averaged spectra in addition to a subset of the time-dependent longitudinally resolved spectra.

#### 3.3.1. Rotationally averaged spectra.
Figure 8 shows the phase-corrected spectral albedo of the modeled planet from 0.4 to $1.4\,\mu m$. The diurnally averaged top-of-atmosphere spectral albedo of the "halophile planet" is up to a factor of 2 times more reflective than the realistic Earth case from 0.55 to $0.85\,\mu m$ and produces a characteristically different spectrum. The halophile Earth spectrum shows an increase in albedo of 25% from a local minimum at $0.5\,\mu m$ to the reflectance peak at $0.68\,\mu m$, just shortward of a water band. The difference in albedo from the 3-D halophile planet spectrum and the 1-D halophile surface spectrum results both from the introduction of clouds and the inclusion of multiple surface types in the 3-D model. The local minimum in the Earth's spectrum at $\sim 0.6\,\mu m$ is the result of absorption from the broad Chappuis ozone band (0.5–$0.7\,\mu m$). However, in the halophile ocean planet spectrum, the minimum is instead seen at $0.5\,\mu m$ due to the combination of increasing reflectivity with wavelength of the halophile pigments on the surface and the atmospheric absorption due to the ozone band.

#### 3.3.2. Longitudinally resolved spectra.
Figure 9 shows the spectral albedo of the disk-averaged "halophile planet" at disk views with subspacecraft longitudes of 160°W and 20°E. These correspond to times when the planet is dominated by the Pacific Ocean (160°W) and the African and parts of the European and Asian continents (20°E). Unsurprisingly, the nonphotosynthetic halophile pigmentation is more detectible when it constitutes a larger fraction of the surface. The halophile-dominated ocean is more reflective at most visible wavelengths than the continents, which is the opposite of modern Earth's spectral contrast between land and ocean. The magnitude of this change at $0.68\,\mu m$ is $\sim 13\%$. In near-infrared wavelengths, the land is more reflective than the halophile-dominated ocean, which is similar to Earth's ocean and land contrast. These effects would combine to produce unique spectral and time-dependent behavior as a function of rotational phase. This heterogeneity would allow the use of time-dependent broadband colors to determine the planetary rotation period and reconstruction of the longitudinal land distribution (Cowan *et al.*, 2009, 2011; Cowan and Strait, 2013).

The experimental reflectance spectra from Section 3.1, the 1-D model synthetic spectra from Section 3.2, and the 3-D model synthetic spectra from Section 3.3 are available on the Virtual Planetary Laboratory's database Web site[3].

### 3.4. Broadband colors

In this subsection, we present and compare broadband colors of pigmented surfaces with and without effects from an overlying atmosphere.

TABLE 5. ONE-DIMENSIONAL SYNTHETIC SPECTRA FEATURE STRENGTHS

| Surface organism | Wavelengths of strongest spectral feature (µm) | Maximum change in reflectance factor ($f_{\Delta R}$) |
|---|---|---|
| Category 1—Strong Spectral Break and Reflective in the Near IR | | |
| B. aurantiacum | 0.52–0.58 | 0.53 |
| C. tepidum | 0.82–0.86 | 1.17 |
| D. radiodurans | 0.54–0.58 | 0.35 |
| H. salinarum | 0.56–0.70 | 2.35 |
| M. luteus | 0.49–0.510 | 0.21 |
| R. radiotolerans | 0.56–0.60 | 0.21 |
| Conifers | 0.69–0.76 | 6.96 |
| Category 2—Gradual Rise into the Near IR | | |
| J. lividum | 0.70–0.80 | 1.88 |
| P. inhibens | — | — |
| S. marcescens | 0.44–0.46 | 0.05 |
| Category 3—Strong Features in Visible and Absorptive in Near IR | | |
| R. sphaeroides | 0.54–0.57 | 0.53 |
| R. palustris | 0.54–0.57 | 0.60 |

The strengths of surface spectral reflectance features from pigmented organisms as seen through an Earth-like atmosphere. See Sections 2.4.1 and 3.2 for a description of the 1-D synthetic spectra.

---

[3]https://depts.washington.edu/naivpl/content/spectral-databases-and-tools

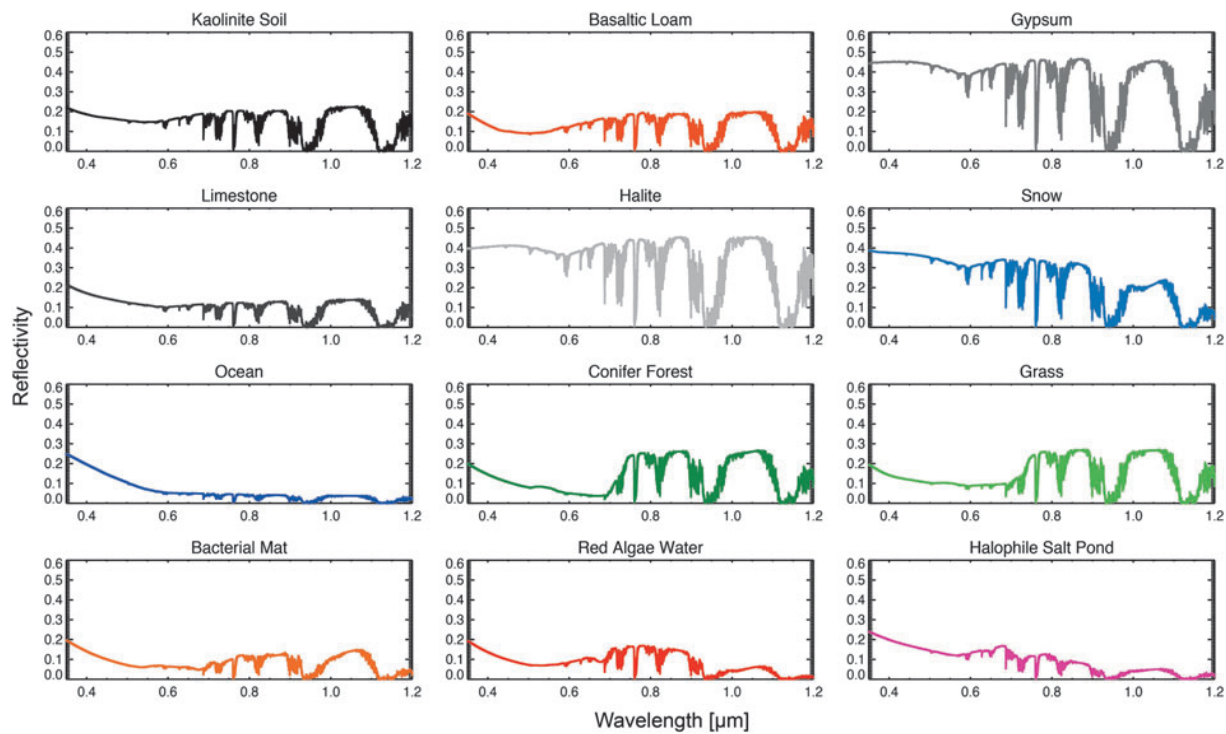

**FIG. 6.** Resulting top-of-atmosphere spectral reflectivity of planets dominated by the abiotic and biotic surfaces described in Section 2.4 and whose spectral reflectances are shown in Fig. 3. The average solar zenith angle is assumed to be 60°. The reflectivity is calculated as the modeled top-of-atmosphere spectral irradiance divided by the spectral irradiance of the Sun and not corrected for solar zenith angle. For the purpose of illustration, the spectra have been smoothed with a moving 5-point statistical mean.

**3.4.1. Broadband colors of pigmented surfaces without atmospheric effects.** A color-color plot of the idealized broadband colors described in Section 2.5 is shown in Fig. 10. These surface-only colors could be considered ''no atmosphere'' scenarios or cases where the effects of the atmosphere have been perfectly removed. The colors occupy a wide range in color-color space with the scatter meeting or exceeding that for abiotic surface types shown here, though seawater, red algae water (from the USGS spectral library), and the conifer forest present clear outliers. The forest presents a stronger red edge than grass or a bacterial mat, giving it a redder B*-I* index.

**3.4.2. Broadband colors of pigmented surfaces as seen through an atmosphere.** Figure 11 shows the broadband colors of abiotic and biological surfaces from the USGS and ASTER spectral libraries, and colors resulting from the normalized spectral reflectivities or albedos of the 1-D and 3-D spectral models. The colors of the 1-D (no clouds) and 3-D (heterogeneous surface and clouds) model spectra include the effects of scattering and absorption in the atmosphere, while the two 3-D model spectral colors additionally include the effects of realistic clouds. As expected, the effect of Rayleigh scattering makes the colors of the clear-sky spectra almost uniformly bluer. Red algae water, basaltic loam, and trees, apparent outliers when considering only the surface reflectance function in the color calculation, begin to populate similar areas in B*-V* versus B*-I* color space when the effects of scattering in a planetary atmosphere are included. The presence of an atmosphere therefore reduces the spread in the distribution of surface colors. This is consistent with work by Sanromá *et al.,* (2013), who included fewer surface types in their analysis but noted a similar trend. For the two cases where heterogeneous surface distributions are assumed and cloud cover is included, that is, the ocean-dominated realistic Earth case and the halophile-dominated oceans case, colors are redder than the 1-D clear-sky cases but still bluer than seen for the airless, pure surface spectra cases. This reddening in comparison to the clear-sky cases is due to the effects of clouds, which truncate the atmospheric column, thereby decreasing Rayleigh scattering, while also behaving as gray scatterers.

**FIG. 7.** Reflectivity spectra of the conifer forest, hypersaline pond, and ocean planet from Fig. 6 are shown together for contrast. The dashed line brackets the wavelength region containing the largest feature seen in the reflectance spectrum of *Halobacterium salinarum*, which contains similar pigments to the biota in the San Francisco salt ponds. The dot-dashed line corresponds to the local maximum reflectivity of the halophile salt pond at ~0.68 μm. For the purpose of illustration, the spectra have been smoothed with a moving 5-point statistical mean.

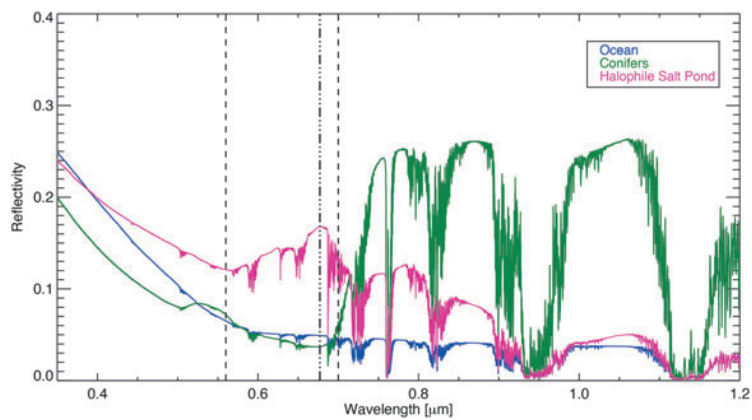

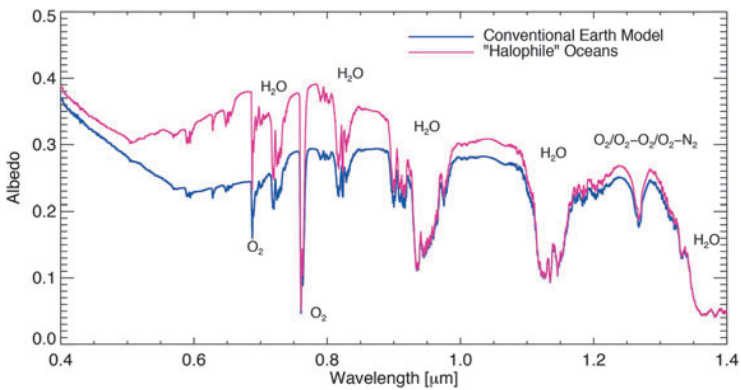

**FIG. 8.** Diurnally averaged spectral albedos from the Virtual Planetary Laboratory's 3-D Earth model for March 18–19, 2008, under a realistic scenario (blue) from Robinson *et al.* (2011) and a scenario that is identical with the exception that the spectral reflectance of the ocean surface has been replaced by the spectral reflectance of a pigmented halophile-dominated saltern pond from work by Dalton *et al.* (2009) (pink). Major gaseous absorption features are labeled. For the purpose of illustration, the spectra have been smoothed with a moving 20-point statistical mean.

In Fig. 12, we show the broadband colors of the simulated planet spectra with the experimentally measured surfaces viewed through an Earth-like atmosphere (from Figs. 5 and 6). Similar to the cases shown for Fig. 11, the colors become bluer with the addition of spectral effects from the atmosphere.

## 4. Discussion

In this study, we explored nonphotosynthetic pigments as potential biosignatures and found that nonphotosynthetic organisms display a rich diversity of spectral features. Within the example set of nonphotosynthetic organisms investigated in this work, significant spectral breaks (> ~10% change in reflectance) are found throughout the visible spectrum from 0.44 to 0.7 μm. In contrast, Kiang *et al.* (2007a) showed that red-edge breaks for photosynthetic vascular plants, lichens, moss, algae, and sea grass clustered between 0.69 and 0.73 μm. Additionally, we find that for some visibly pigmented organisms, such as *Phaeobacter inhibens*, no significant spectral-break features are present.

### 4.1. Challenges in identifying biological reflectance spectra signatures

We have also shown that pigmented organisms do not occupy a unique region of visible broadband color-color space and that the broadband colors of biological and abiotic surfaces are altered significantly when viewed through a planetary atmosphere. Previous work has been done to calculate the visible broadband colors of biotic and abiotic surfaces, but that work did not include the effects of an overlying atmosphere (Hegde and Kaltenegger, 2013). The atmosphere's effect on the observed colors is due primarily to Rayleigh scattering, which drives colors to bluer regions of color-color space, although absorption from water vapor and ozone also modifies the colors inferred from the surface alone. A planet dominated by even a single surface type could exhibit a variety of broadband colors depending on the overlying atmosphere's level of Rayleigh scattering and cloud cover. Rayleigh scattering is modulated by the thickness of the atmospheric column and may be difficult to determine, as some atmospheric and surface absorbing species can mask the presence of Rayleigh scattering even for planets with substantial atmospheres (Crow *et al.,* 2010). Realistically, planets will have a variety of surface types, atmospheric compositions and masses, and cloud composition and fractional coverage. The observed colors will therefore be highly degenerate, with similar colors being produced by a large variety of possible planetary environments.

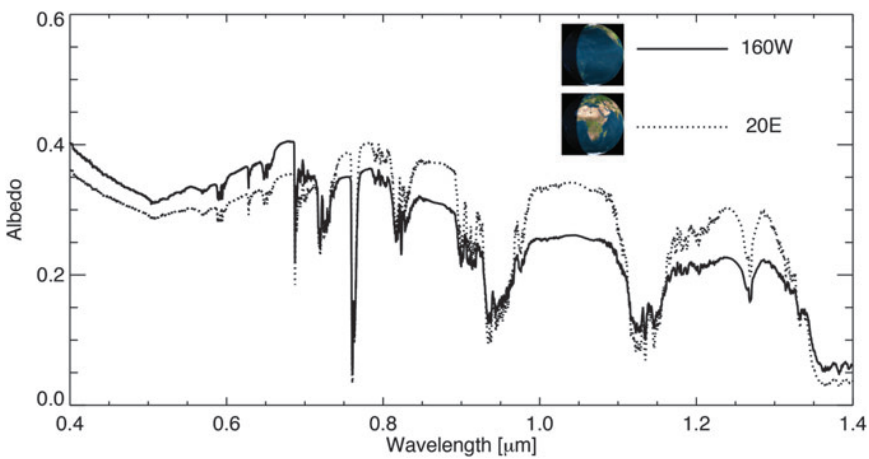

**FIG. 9.** Spectra from the halophile ocean version of the Earth model at two discrete sub-observer longitudes: 160°W (over the Pacific Ocean) and 20°E (centered over the African continent and Eurasia). Note the difference in the reflectance peak at ~0.68 μm produced by the halophiles' carotenoid pigments suspended in water. Insets were generated with the Earth and Moon Viewer (http://www.fourmilab.ch/earthview). For the purpose of illustration, the spectra have been smoothed with a moving 20-point statistical mean.

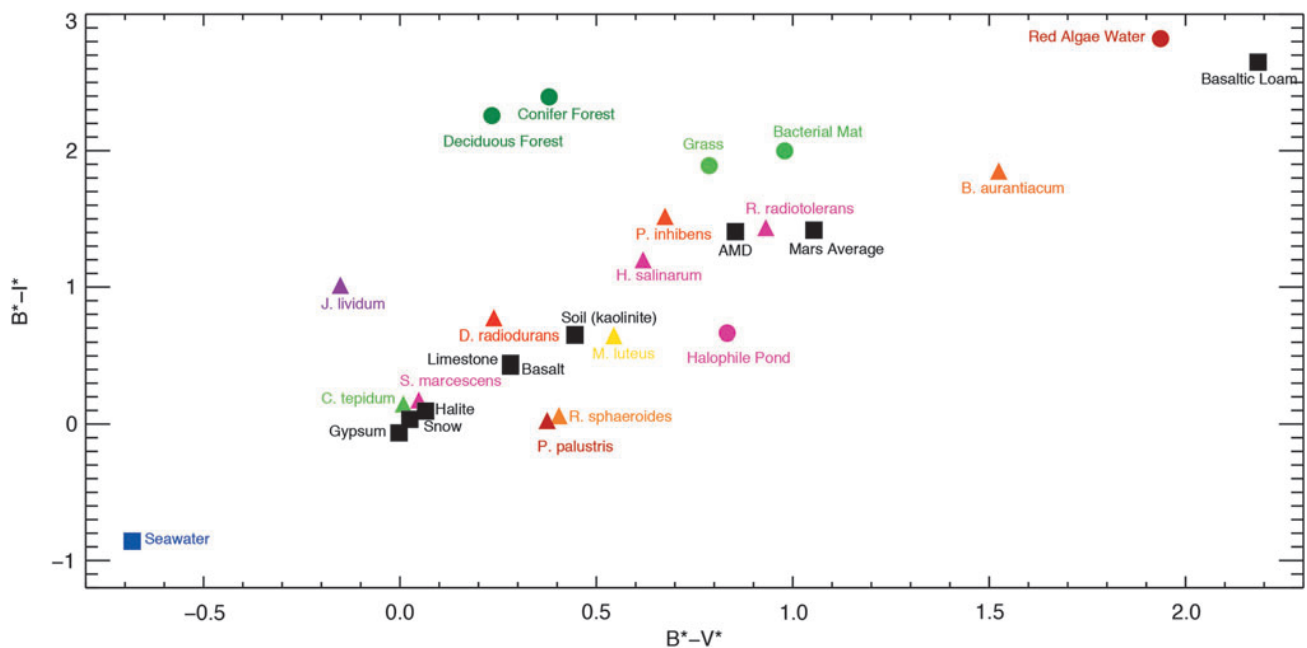

**FIG. 10.** Color-color plot of pigmented microorganisms from this work and biotic and abiotic surfaces from spectral databases. The method to derive colors is described in Section 2.4. Triangles represent colors of organisms derived from reflectance spectra measured for this work; squares are abiotic reflectances from spectra taken from the USGS spectral library (Clark *et al.*, 2007); circles are biotic reflectances from the USGS spectral library or the ASTER spectral library (Baldridge *et al.*, 2009).

The surface reflection signal of any possible surface biota will also be a component of a larger, complex planetary environment, which will have its own effects on the re-flectance spectrum. This will make attributing specific spectral features to life implausible without information about the planetary context. If the dominant pigments for the organisms are nonphotosynthetic, there is a large range of possible wavelengths where significant spectral features may appear. Analogues of the VRE may be predicted based on the wavelength for the peak photon-flux at the planetary surface if the incident stellar spectrum and the planetary atmospheric parameters are well known (Kiang *et al.*, 2007a, 2007b). However, if the most widespread or detectable surface organisms are not photosynthetic, then the strongest spectral breaks may not be at the predicted wavelengths. Even many anoxygenic photosynthesizers on Earth do not have spectral breaks near 0.7 μm due to bacteriochlorophylls that absorb at longer wavelengths (Hohmann-Marriott and Blankenship, 2012). For this reason, and due to possible degeneracies with other planetary properties, a few preselected narrowband filters would not be able to distinguish between the vast diversity of possible features, and the best discrimination of the source of the surface reflectivity would likely require an evaluation of the whole visible spectrum with spectrally resolved data.

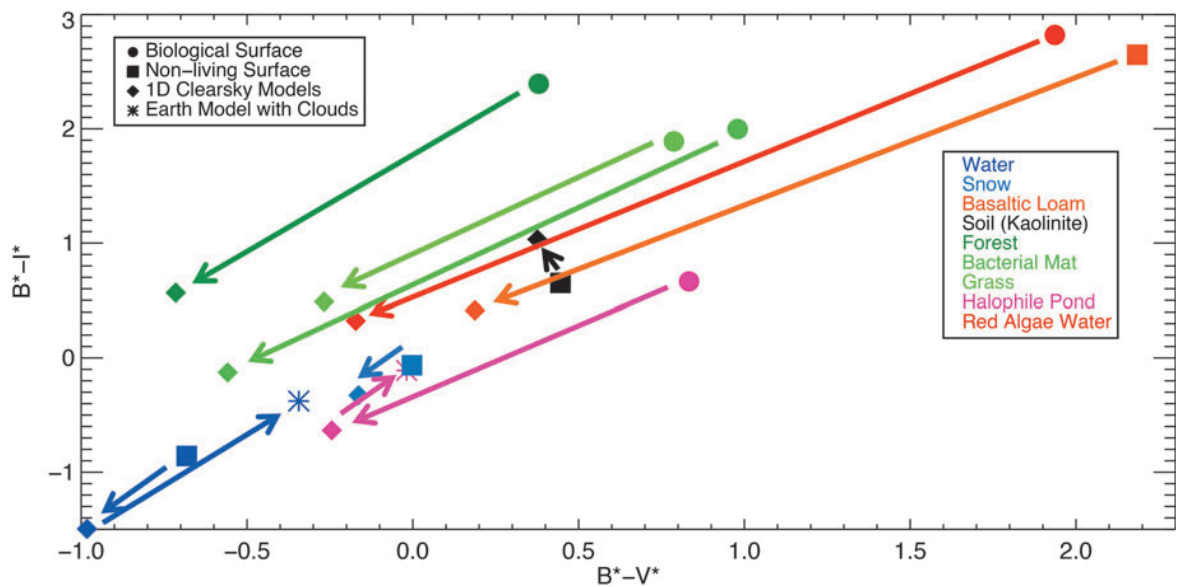

**FIG. 11.** Color-color plot comparing the normalized colors of lifeless surfaces (squares), pigmented biological surfaces (circles), these surfaces under an Earth-like cloudless atmosphere (diamonds), and two cases that are diurnal averages of a heterogeneous planet with realistic clouds (asterisks). Each set is color-coded in the figure legend. Arrows are drawn from the colors with no radiative effects from the atmosphere included, to the colors including effects from the atmosphere.

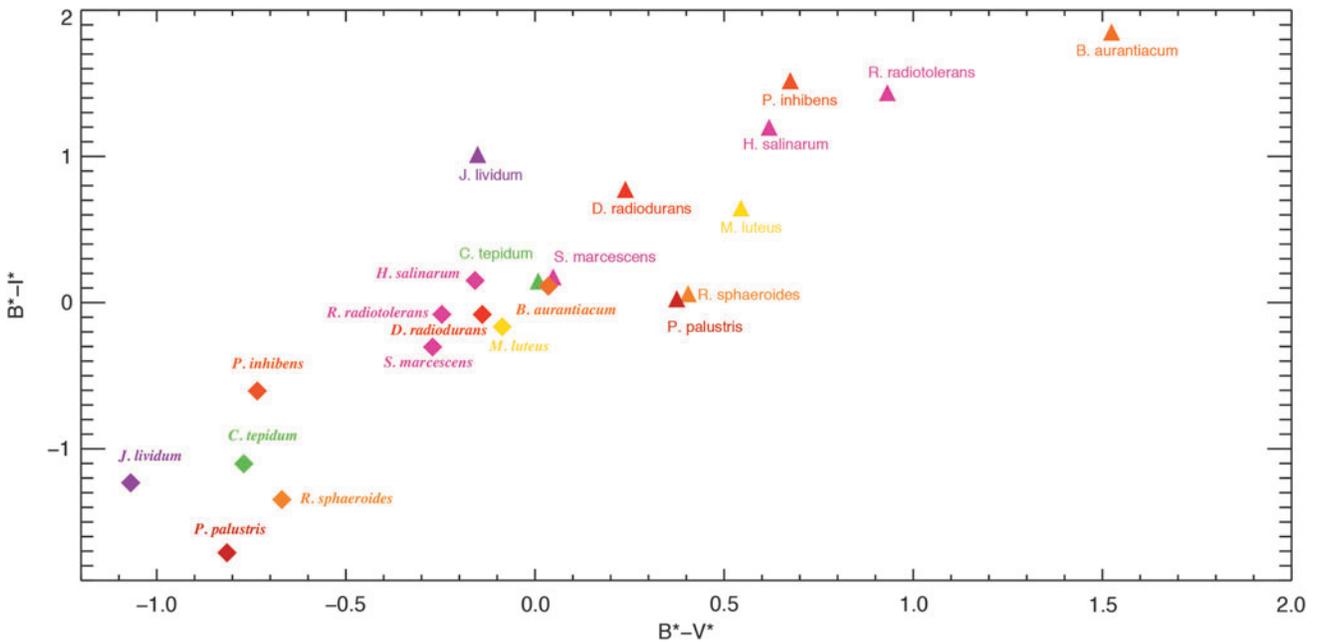

**FIG. 12.** Color-color plot comparing the normalized colors of experimentally measured microbial spectral reflectances (triangles, non-italicized text labels) and those of surfaces dominated by these pigmented organisms under an Earth-like cloudless atmosphere (diamonds, italicized text labels). Note that the colors for pigmented microbial surfaces shown here are the same for Fig. 10, but the axis limits are different to accommodate the spectral model colors.

Examination of the planetary context could also rule out some of the false-positive possibilities for biological surface reflectance features. For example, the minerals, such as sulfur compounds, that have been posited as a possible false positive for the VRE (Seager *et al.*, 2005) are much brighter at near-infrared wavelengths than water (Clark *et al.*, 2007; Baldridge *et al.*, 2009). Therefore, if a larger spectral range is accessible, it may be possible to discriminate minerals from pigments suspended in water, as the latter will show much lower albedos at near-infrared wavelengths because of absorption from the overlying water.

### 4.2. Plausibility of widespread environments conducive to nonphotosynthetic life

Pigmented nonphotosynthetic organisms inhabit environments on Earth that could exist on other habitable terrestrial planets. An example of this type of environment is hypersaline ponds, whose spectral reflectances are often dominated by nonphotosynthetic pigmented halophiles. The ''halophile planet'' modeled in this study is a best-case scenario for a planet that possesses significant coverage of nonphotosynthetic pigments, although it is plausible that similar environments may exist on terrestrial planets with the right conditions. For example, high-salinity environments may be more extensive on a planet with shallower oceans or different weathering rates (Pierrehumbert, 2010). The high-salinity environment is hostile to many potential grazers, and the resulting lack of predation allows the microbial inhabitants of high-salt ponds to exist at high densities even with low levels of primary productivity (Oren, 2009). The same type of scenario is observed in high-temperature hot springs, where thermophilic bacteria can grow into biofilms in part because they face little to no predation. Other ''extreme'' environmental conditions, such as an anoxic atmosphere, may also preclude macroscopic grazers, as was the case on Earth before the origin of animals ∼575 million years ago (Narbonne, 2005). The suppression of predation due to lack of oxygen could potentially allow for a much larger—and therefore more detectable—steady-state biomass, without requiring extremes in salinity or temperature.

### 4.3. Remote detectability of nonphotosynthetic life with spectrally resolved data

In contrast to information from broadband filters, spectrally resolved data is sensitive to reflectance features, such as spectral breaks, that may be indicative of surface life, such as photosynthetic analogues to the VRE or the nonphotosynthetic organisms included in this study. The reflective pigments of nonphotosynthetic organisms suspended in water may be distinguished from abiotic features by examining the spectral shape of the feature and determining the environmental context from the full spectrum; for example, by detecting strong water vapor or searching for ocean glint (Williams and Gaidos, 2008; Robinson *et al.*, 2010, 2014). Glint that is spatially correlated with the observed pigment signature (using a mapping technique such as that described in Cowan *et al.*, 2009) would argue for colocation on the planetary surface.

For an environment hosting widespread nonphotosynthetic pigmented halophilic organisms, our 3-D spectral model predicts a characteristic spectral feature, and a time-varying signal in the disk-averaged spectrum that is potentially several times larger than that posited for the VRE. The VRE signal on Earth has been quantified from Earthshine data as the time-dependent fractional change in albedo at 0.7 $\mu$m due to the changing proportion of vegetation in the disk view of the planet (Montanes-Rodriguez *et al.*, 2006). The VRE signal defined in this way is a ~2% effect, though this quantification is dependent on the viewing geometry of the planet and would have varied through geological time (Arnold *et al.*, 2009). In contrast, a shallow ocean populated by a significant density of pigmented halophiles could produce a maximum 13% time-variable signal at 0.68 $\mu$m, assuming the modern Earth's continental arrangement. The photometric variability due to changes in the proportional coverage of the visible planetary disk has been considered in other model scenarios as a method to distinguish biological surfaces from abiotic ones (Sanromá *et al.*, 2013, 2014). The increasing variability with wavelength of the amplitude of the infrared albedo modeled for the Archean Earth with different fractional coverage of anoxygenic photosynthetic purple bacteria (Sanromá *et al.*, 2014) is due to the high reflectivity of purple bacteria at near-infrared wavelengths (~1.1 $\mu$m). In contrast, the pigmented organisms found in the salt ponds are more reflective at visible wavelengths, and a planet populated with them would instead exhibit the greatest difference in spectral albedo between ocean and land in the visible portion of the spectrum.

### 4.4. Near-future implications

Near-future space-based direct-imaging telescopes are currently being designed that can probe the spectral reflectance of planets in the habitable zones of their stars. However, due to their relatively small apertures ($\leq$ 1.4 m), these telescopes will not be able to observe wavelengths as long as those of the photosynthetic red-edge feature for planets in the habitable zones of all but the nearest stars (Seager *et al.*, 2014; Stapelfeldt *et al.*, 2014). This results from the relationship between telescope diameter, inner working angle, and spectral range. Even for instruments with spectral ranges that extend over the entire astronomical visible wavelength range (0.4–1.0 $\mu$m), a significant portion of range will be excluded for many targets depending on their distance from Earth and consequently whether their habitable zones are accessible given the inner working angle requirements for those wavelengths. In the case of long wavelength cutoffs shorter than 0.7 $\mu$m, which also excludes the strongest $O_2$ and $CH_4$ spectral features, surface reflectance biosignatures shortward of the traditional VRE (such as those shown in this paper) may be the only plausibly accessible class of biosignature.

### 5. Summary and Conclusions

With laboratory measurements of wavelength-dependent reflectance, we have demonstrated the diversity of spectral features and broadband colors of organisms with nonphotosynthetic pigments. Using radiative transfer models of 1-D and 3-D, planet-wide, spatially resolved spectra, we have shown that, given adequate surface coverage, a nonphotosynthetic pigment can create a significant effect in the disk-averaged spectrum of a terrestrial planet. We also calculated broadband colors for a range of pigments and other biological and abiotic surface types in the visible wavelength range considering only their spectral reflectances, and for more realistic scenarios that included an overlying atmosphere and clouds. While the resultant broadband colors are dependent on the nature of spectral reflectance of the surface—including those of any biologically produced pigments—it would not be feasible to make a definitive determination of surface type based on the broadband color alone. Higher-resolution spectra would be far more informative, allowing observers to search for features such as line breaks or peaks that may indicate biological pigments. Highly reflective pigments suspended in water would possess a spectral reflectance that peaks in the visible spectrum and would exhibit glint at small phase angles, possibly allowing an observer to distinguish between a nonphotosynthetic pigment biosignature and another reflective surface. This work has shown that there is a diverse range of potential surface reflectance biosignatures beyond the commonly considered photosynthetic red edge. For many of the planned first-generation terrestrial exoplanet direct-imaging observations, spectral features from nonphotosynthetic pigments may be the only plausibly accessible biosignatures in the visible wavelength ranges where data may be obtained.

### Acknowledgments

This work was supported in part by the NASA Astrobiology Institute's Virtual Planetary Laboratory Lead Team, funded by the National Aeronautics and Space Administration through the NASA Astrobiology Institute under solicitation NNH12ZDA002C and Cooperative Agreement Number NNA13AA93A. This work was also supported in part by the UK Centre for Astrobiology.

We would like to thank Nancy Kiang and Niki Parenteau for helpful comments and discussion. We would also like to thank the two anonymous reviewers who provided useful comments that allowed us to greatly improve this manuscript.

### Author Disclosure Statement

No competing financial interests exist.

### References

Archetti, M., Döring, T.F., Hagen, S.B., Hughes, N.M., Leather, S.R., Lee, D.W., Lev-Yadun, S., Manetas, Y., Ougham, H.J., Schaberg, P.G., and Thomas, H. (2009) Unravelling the evolution of autumn colours: an interdisciplinary approach. *Trends Ecol Evol* 24:166–173.

Arnold, L., Bréon, F.-M., and Brewer, S. (2009) The Earth as an extrasolar planet: the vegetation spectral signature today and during the last Quaternary climatic extrema. *International Journal of Astrobiology* 8:81–94.

Baldridge, A.M., Hook, S.J., Grove, C.I., and Rivera, G. (2009) The ASTER spectral library version 2.0. *Remote Sens Environ* 113:711–715.

Bennett, J.W. and Bentley, R. (2000) Seeing red: the story of prodigiosin. *Adv Appl Microbiol* 47:1–32.

Boichenko, V.A., Wang, J.M., Antón, J., Lanyi, J.K., and Balashov, S.P. (2006) Functions of carotenoids in xanthorhodopsin and archaerhodopsin, from action spectra of photoinhibition of cell respiration. *Biochim Biophys Acta* 1757:1649–1656.

Bosak, T., Greene, S.E., and Newman, D.K. (2007) A likely role for anoxygenic photosynthetic microbes in the formation of ancient stromatolites. *Geobiology* 5:119–126.

Brock, T.D. and Freeze, H. (1969) *Thermus aquaticus* gen. n. and sp. n., a non-sporulating extreme thermophile. *J Bacteriol* 98:289–297.

Broschat, S.L., Loge, F.J., Peppin, J.D., White, D., Call, D.R., and Kuhn, E. (2014) Optical reflectance assay for the detection of biofilm formation. *J Biomed Opt* 10, doi:10.1117/1.1953347.

Buick, R. (2008) When did oxygenic photosynthesis evolve? *Philos Trans R Soc Lond B Biol Sci* 363:2731–2743.

Cao, C., Love, G.D., Hays, L.E., Wang, W., Shen, S., and Summons, R.E. (2009) Biogeochemical evidence for euxinic oceans and ecological disturbance presaging the end-Permian mass extinction event. *Earth Planet Sci Lett* 281:188–201.

Chen, M.-Y., Wu, S.-H., Lin, G.-H., Lu, C.-P., Lin, Y.-T., Chang, W.-C., and Tsay, S.-S. (2004) *Rubrobacter taiwanensis* sp. nov., a novel thermophilic, radiation-resistant species isolated from hot springs. *Int J Syst Evol Microbiol* 54:1849–1855.

Chittka, L. and Raine, N.E. (2006) Recognition of flowers by pollinators. *Curr Opin Plant Biol* 9:428–435.

Clark, R.N., Swayze, G.A., Wise, R., Livo, E., Hoefen, T., Kokaly, R., and Sutley, S.J. (2007) *USGS Digital Spectral Library splib06a*, Digital Data Series 231, U.S. Geological Survey, Reston, VA. Available online at http://speclab.cr.usgs.gov/spectral.lib06.

Clayton, R.K. (1966) Spectroscopic analysis of bacteriochlorophylls *in vitro* and *in vivo*. *Photochem Photobiol* 5:669–677.

Cockell, C.S., Léger, A., Fridlund, M., Herbst, T.M., Kaltenegger, L., Absil, O., Beichman, C., Benz, W., Blanc, M., Brack, A., Chelli, A., Colangeli, L., Cottin, H., Coudé du Foresto, F., Danchi, W.C., Defrère, D., den Herder, J.-W., Eiroa, C., Greaves, J., Henning, T., Johnston, K.J., Jones, H., Labadie, L., Lammer, H., Launhardt, R., Lawson, P., Lay, O.P., LeDuigou, J.M., Liseau, R., Malbet, F., Martin, S.R., Mawet, D., Mourard, D., Moutou, C., Mugnier, L.M., Ollivier, M., Paresce, F., Quirrenbach, A., Rabbia, Y.D., Raven, J.A., Rottgering, H.J., Rouan, D., Santos, N.C., Selsis, F., Serabyn, E., Shibai, H., Tamura, M., Thiébaut, E., Westall, F., and White, G.J. (2009) Darwin—a mission to detect and search for life on extrasolar planets. *Astrobiology* 9:1–22.

Cogdell, R.J., Howard, T.D., Bittl, R., Schlodder, E., Geisenheimer, I., and Lubitz, W. (2000) How carotenoids protect bacterial photosynthesis. *Philos Trans R Soc Lond B Biol Sci* 355:1345–1349.

Cowan, N.B. and Strait, T.E. (2013) Determining reflectance spectra of surfaces and clouds on exoplanets. *Astrophys J* 765, doi:10.1088/2041-8205/765/1/L17.

Cowan, N.B., Agol, E., Meadows, V.S., Robinson, T., Livengood, T.A., Deming, D., Lisse, C.M., A'Hearn, M.F., Wellnitz, D.D., Seager, S., and Charbonneau, D. (2009) Alien maps of an ocean-bearing world. *Astrophys J* 700:915–923.

Cowan, N.B., Robinson, T., Livengood, T.A., Deming, D., Agol, E., A'Hearn, M.F., Charbonneau, D., Lisse, C.M., Meadows, V.S., Seager, S., Shields, A.L., and Wellnitz, D.D. (2011) Rotational variability of Earth's polar regions: implications for detecting snowball planets. *Astrophys J* 731, doi:10.1088/0004-637X/731/1/76.

Cox, M.M. and Battista, J.R. (2005) *Deinococcus radiodurans*—the consummate survivor. *Nat Rev Microbiol* 3:882–892.

Crisp, D. (1997) Absorption of sunlight by water vapor in cloudy conditions: a partial explanation for the cloud absorption anomaly. *Geophys Res Lett* 24:571–574.

Crow, C.A., McFadden, L.A., Robinson, T., Livengood, T.A., Hewagama, T., Barry, R.K., Deming, L.D., Meadows, V., and Lisse, C.M. (2010) Views from EPOXI: colors in our solar system as an analog for extrasolar planets. *Astrophys J* 729, doi:10.1088/0004-637X/729/2/130.

Dadachova, E., Bryan, R.A., Huang, X., Moadel, T., Schweitzer, A.D., Aisen, P., Nosanchuk, J.D., and Casadevall, A. (2007) Ionizing radiation changes the electronic properties of melanin and enhances the growth of melanized fungi. *PloS One* 2, doi:10.1371/journal.pone.0000457.

Dalton, J.B., Mogul, R., Kagawa, H.K., Chan, S.L., and Jamieson, C.S. (2003) Near-infrared detection of potential evidence for microscopic organisms on Europa. *Astrobiology* 3:505–529.

Dalton, J.B., Palmer-Moloney, L.J., Rogoff, D., Hlavka, C., and Duncan, C. (2009) Remote monitoring of hypersaline environments in San Francisco Bay, CA, USA. *Int J Remote Sens* 30:2933–2949.

Dartnell, L. (2011) Biological constraints on habitability. *Astronomy & Geophysics* 52:25–28.

DasSarma, S. (2006) Extreme halophiles are models for astrobiology. *Microbe Wash DC* 1:120–126.

Decho, A.W., Kawaguchi, T., Allison, M.A., Louchard, E.M., Reid, R.P., Stephens, F.C., Voss, K.J., Wheatcroft, R.A., and Taylor, B.B. (2003) Sediment properties influencing upwelling spectral reflectance signatures: the "biofilm gel effect." *Limnol Oceanogr* 48:431–443.

Des Marais, D.J., Harwit, M.O., Jucks, K.W., Kasting, J.F., Lin, D.N.C., Lunine, J.I., Schneider, J., Seager, S., Traub, W.A., and Woolf, N.J. (2002) Remote sensing of planetary properties and biosignatures on extrasolar terrestrial planets. *Astrobiology* 2:153–181.

Des Marais, D.J., Nuth, J.A., Allamandola, L.J., Boss, A.P., Farmer, J.D., Hoehler, T.M., Jakosky, B.M., Meadows, V.S., Pohorille, A., Runnegar, B., and Spormann, A.M. (2008) The NASA Astrobiology Roadmap. *Astrobiology* 8:715–730.

Dogs, M., Voget, S., Teshima, H., Petersen, J., Davenport, K., Dalingault, H., Chen, A., Pati, A., Ivanova, N., Goodwin, L.A., Chain, P., Detter, J.C., Standfest, S., Rohde, M., Gronow, S., Kyrpides, N.C., Woyke, T., Simon, M., Klenk, H., Goker, M., and Brinkoff, T. (2013) Genome sequence of *Phaeobacter inhibens* type strain (T5(T)), a secondary metabolite producing representative of the marine Roseobacter clade, and emendation of the species description of *Phaeobacter inhibens*. *Stand Genomic Sci* 9:334–350.

Domagal-Goldman, S.D., Segura, A., Claire, M.W., Robinson, T.D., and Meadows, V.S. (2014) Abiotic ozone and oxygen in atmospheres similar to prebiotic Earth. *Astrophys J* 792, doi:10.1088/0004-637X/792/2/90.

Durán, N., Justo, G.Z., Ferreira, C.V., Melo, P.S., Cordi, L., and Martins, D. (2007) Violacein: properties and biological activities. *Biotechnol Appl Biochem* 48:127–133.

Gavrish, E.I., Krauzova, V.I., Potekhina, N.V., Karasev, S.G., Plotnikova, E.G., Altyntseva, O.V, Korosteleva, L.A., and Evtushenko, L.I. (2004) [Three new species of brevibacteria—*Brevibacterium antiquum* sp. nov., *Brevibacterium auriantiacum* sp. nov. and *Brevibacterium permense* sp. nov]. In Russian. *Mikrobiologiia* 73:218–225.

Glaeser, J. and Klug, G. (2005) Photo-oxidative stress in *Rhodobacter sphaeroides*: protective role of carotenoids and expression of selected genes. *Microbiology* 151:1927–1938.

González, J.E. and Keshavan, N.D. (2006) Messing with bacterial quorum sensing. *Microbiol Mol Biol Rev* 70:859–875.

Gorski, K.M., Hivon, E., Banday, A.J., Wandelt, B.D., Hansen, F.K., Reinecke, M., and Bartelman, M. (2004) HEALPix—a framework for high resolution discretization, and fast analysis of data distributed on the sphere. *Astrophys J* 622:759–771.

Greenblatt, C.L., Baum, J., Klein, B.Y., Nachshon, S., Koltunov, V., and Cano, R.J. (2004) *Micrococcus luteus*—survival in amber. *Microb Ecol* 48:120–127.

Grote, M. and O'Malley, M.A. (2011) Enlightening the life sciences: the history of halobacterial and microbial rhodopsin research. *FEMS Microbiol Rev* 35:1082–1099.

Haddix, P.L., Jones, S., Patel, P., Burnham, S., Knights, K., Powell, J.N., and LaForm, A. (2008) Kinetic analysis of growth rate, ATP, and pigmentation suggests an energy-spilling function for the pigment prodigiosin of *Serratia marcescens. J Bacteriol* 190:7453–7463.

Haddock, S.H.D., Moline, M.A., and Case, J.F. (2010) Bioluminescence in the sea. *Ann Rev Mar Sci* 2:443–493.

Hegde, S. and Kaltenegger, L. (2013) Colors of extreme exo-Earth environments. *Astrobiology* 13:47–56.

Hejazi, A. and Falkiner, F.R. (1997) *Serratia marcescens. J Med Microbiol* 46:903–912.

Hitchcock, D.R. and Lovelock, J.E. (1967) Life detection by atmospheric analysis. *Icarus* 7:149–159.

Hohmann-Marriott, M.F. and Blankenship, R.E. (2012) The photosynthetic world. In *Photosynthesis: Plastid Biology, Energy Conversion and Carbon Assimilation*, edited by J.J. Eaton-Rye, B.C. Tripathy, and T.D. Sharkey, Springer, Dordrecht, the Netherlands, pp 3–32.

Jorgensen, B.B. and Des Marais, D.J. (1988) Optical properties of benthic photosynthetic communities: fiber-optic studies of cyanobacterial mats. *Limnol Oceanogr* 33:99–113.

Keeling, C.D. (1960) The concentration and isotopic abundances of carbon dioxide in the atmosphere. *Tellus* 12:200–203.

Keeling, C.D., Chin, J.F.S., and Whorf, T.P. (1996) Increased activity of northern vegetation inferred from atmospheric $CO_2$ measurements. *Nature* 382:146–149.

Khalil, M.A.K. and Rasmussen, R.A. (1983) Sources, sinks, and seasonal cycles of atmospheric methane. *J Geophys Res* 88, doi:10.1029/JC088iC09p05131.

Kiang, N.Y., Siefert, J., Govindjee, and Blankenship, R.E. (2007a) Spectral signatures of photosynthesis. I. Review of Earth organisms. *Astrobiology* 7:222–251.

Kiang, N.Y., Segura, A., Tinetti, G., Govindjee, Blankenship, R.E., Cohen, M., Siefert, J., Crisp, D., and Meadows, V.S. (2007b) Spectral signatures of photosynthesis. II. Coevolution with other stars and the atmosphere on extrasolar worlds. *Astrobiology* 7:252–274.

Kimmel, K.E. and Maier, S. (1969) Effect of cultural conditions on the synthesis of violacein in mesophilic and psychrophilic strains of Chromobacterium. *Can J Microbiol* 15:111–116.

Kimura, H., Asada, R., Masta, A., and Naganuma, T. (2003) Distribution of microorganisms in the subsurface of the Manus Basin hydrothermal vent field in Papua New Guinea. *Appl Environ Microbiol* 69:644–648.

Klassen, J.L. (2010) Phylogenetic and evolutionary patterns in microbial carotenoid biosynthesis are revealed by comparative genomics. *PLoS One* 5, doi:10.1371/journal.pone.0011257.

Kühl, M., Lassen, C., and Revsbech, N.P. (1997) A simple light meter for measurements of PAR (400 to 700 nm) with fiber-optic microprobes: application for P vs E0(PAR) measurements in a microbial mat. *Aquat Microb Ecol* 13:197–207.

Larimer, F.W., Chain, P., Hauser, L., Lamerdin, J., Malfatti, S., Do, L., Land, M.L., Pelletier, D.A., Beatty, J.T., Lang, A.S., Tabita, F.R., Gibson, J.L., Hanson, T.E., Bobst, C., Torres, J.L., Peres, C., Harrison, F.H., Gibson, J., and Harwood, C.S. (2004) Complete genome sequence of the metabolically versatile photosynthetic bacterium *Rhodopseudomonas palustris. Nat Biotechnol* 22:55–61.

Lemee, L., Peuchant, E., Clerc, M., Brunner, M., and Pfander, H. (1997) Deinoxanthin: a new carotenoid isolated from *Deinococcus radiodurans. Tetrahedron* 53:919–926.

Levine, M., Lisman, D., Shaklan, S., Kasting, J., Traub, W., Alexander, J., Angel, R., Blaurock, C., Brown, M., Brown, R., Burrows, C., Clampin, M., Cohen, E., Content, D., Dewell, L., Dumont, P., Egerman, R., Ferguson, H., Ford, V., Greene, J., Guyon, O., Hammel, H., Heap, S., Ho, T., Horner, S., Hunyadi, S., Irish, S., Jackson, C., Kasdin, J., Kissil, A., Krim, M., Kuchner, M., Kwack, E., Lillie, C., Lin, D., Liu, A., Marchen, L., Marley, M., Meadows, V., Mosier, G., Mouroulis, P., Noecker, M., Ohl, R., Oppenheimer, B., Pitman, J., Ridgway, S., Sabatke, E., Seager, S., Shao, M., Smith, A., Soummer, R., Stapelfeldt, K., Tenerell, D., Trauger, J., and Vanderbei, J. (2009) Terrestrial Planet Finder Coronagraph (TPF-C) flight baseline concept. arXiv:0911.3200.

Liu, G.Y. and Nizet, V. (2009) Color me bad: microbial pigments as virulence factors. *Trends Microbiol* 17:406–413.

Livengood, T.A., Deming, L.D., A'Hearn, M.F., Charbonneau, D., Hewagama, T., Lisse, C.M., McFadden, L.A., Meadows, V.S., Robinson, T.D., Seager, S., and Wellnitz, D.D. (2011) Properties of an Earth-like planet orbiting a Sun-like star: Earth observed by the EPOXI mission. *Astrobiology* 11:907–930.

Lovelock, J.E. (1965) A physical basis for life detection experiments. *Nature* 207:568–570.

Luger, R. and Barnes, R. (2015) Extreme water loss and abiotic $O_2$ buildup on planets throughout the habitable zones of M dwarfs. *Astrobiology* 15:119–143.

Mackenzie, C., Eraso, J.M., Choudhary, M., Roh, J.H., Zeng, X., Bruscella, P., Puskás, A., and Kaplan, S. (2007) Postgenomic adventures with *Rhodobacter sphaeroides. Annu Rev Microbiol* 61:283–307.

Martens, T., Heidorn, T., Pukall, R., Simon, M., Tindall, B.J., and Brinkhoff, T. (2006) Reclassification of *Roseobacter gallaeciensis* Ruiz-Ponte *et al.* 1998 as *Phaeobacter gallaeciensis* gen. nov., comb. nov., description of *Phaeobacter inhibens* sp. nov., reclassification of *Ruegeria algicola* (Lafay *et al.* 1995) Uchino *et al.* 1999 as *Marinovum algicola* gen. nov., comb. nov., and emended descriptions of the genera *Roseobacter, Ruegeria* and *Leisingera. Int J Syst Evol Microbiol* 56:1293–1304.

McClean, K.H., Winson, M.K., Fish, L., Taylor, A., Chhabra, S.R., Camara, M., Daykin, M., Lamb, J.H., Swift, S., Bycroft, B.W., Stewart, G.S.A.B., and Williams, P. (1997) Quorum sensing and *Chromobacterium violaceum*: exploitation of violacein production and inhibition for the detection of N-acylhomoserine lactones. *Microbiology* 143:3703–3711.

Meadows, V.S. (2006) Modelling the diversity of extrasolar terrestrial planets. *Proceedings of the International Astronomical Union* 1, doi:10.1017/S1743921306009033.

Meadows, V.S. and Crisp, D. (1996) Ground-based near-infrared observations of the Venus nightside: the thermal structure and water abundance near the surface. *J Geophys Res* 101:4595–4622.

Meeks, J.C. and Castenholz, R.W. (1971) Growth and photosynthesis in an extreme thermophile, *Synechococcus lividus* (Cyanophyta). *Arch Mikrobiol* 78:25–41.

Meyer, J.-M. (2000) Pyoverdines: pigments, siderophores and potential taxonomic markers of fluorescent *Pseudomonas* species. *Arch Microbiol* 174:135–142.

Montanes-Rodriguez, P., Palle, E., Goode, P.R., and Martin-Torres, F.J. (2006) Vegetation signature in the observed globally integrated spectrum of Earth considering simultaneous cloud data: applications for extrasolar planets. *Astrophys J* 651:544–552.

Narbonne, G.M. (2005) The Ediacara biota: Neoproterozoic origin of animals and their ecosystems. *Annu Rev Earth Planet Sci* 33:421–442.

Oren, A. (2009) Saltern evaporation ponds as model systems for the study of primary production processes under hypersaline conditions. *Aquat Microb Ecol* 56:193–204.

Oren, A. (2013) Salinibacter: an extremely halophilic bacterium with archaeal properties. *FEMS Microbiol Lett* 342:1–9.

Oren, A. and Dubinsky, Z. (1994) On the red coloration of saltern crystallizer ponds. II. Additional evidence for the contribution of halobacterial pigments. *International Journal of Salt Lake Research* 3:9–13.

Oren, A., Stambler, N., and Dubinsky, Z. (1992) On the red coloration of saltern crystallizer ponds. *International Journal of Salt Lake Research* 1:77–89.

Painter, T.H., Duval, B., Thomas, W.H., Mendez, M., Heintzelman, S., and Dozier, J. (2001) Detection and quantification of snow algae with an airborne imaging spectrometer. *Appl Environ Microbiol* 67:5267–5272.

Pantanella, F., Berlutti, F., Passariello, C., Sarli, S., Morea, C., and Schippa, S. (2007) Violacein and biofilm production in *Janthinobacterium lividum. J Appl Microbiol* 102:992–999.

Pfennig, N. and Truper, H.G. (1971) Type and neotype strains of the species of phototrophic bacteria maintained in pure culture. *Int J Syst Bacteriol* 21:19–24.

Pierrehumbert, R.T. (2010) *Principles of Planetary Climate,* Cambridge University Press, Cambridge, UK.

Porsby, C.H., Nielsen, K.F., and Gram, L. (2008) *Phaeobacter* and *Ruegeria* species of the *Roseobacter* clade colonize separate niches in a Danish Turbot (*Scophthalmus maximus*)–rearing farm and antagonize *Vibrio anguillarum* under different growth conditions. *Appl Environ Microbiol* 74:7356–7364.

Postman, M., Traub, W.A., Krist, J., Stapelfeldt, K., Brown, R., Oegerle, W., Lo, A., Clampin, M., Soummer, R., Wiseman, J., and Mountain, M. (2010) Advanced Technology Large-Aperture Space Telescope (ATLAST): characterizing habitable worlds. In *Pathways Towards Habitable Planets: Proceedings of a Workshop Held 14 to 18 September 2009 in Barcelona, Spain,* edited by V. Coudé du Foresto, D.M. Gelino, and I. Ribas, Astronomical Society of the Pacific, San Francisco, p 361.

Prescott, L., Harley, J., and Klein, D. (2005) *Microbiology,* McGraw Hill, New York.

Proteau, P.J., Gerwick, W.H., Garcia-Pichel, F., and Castenholz, R. (1993) The structure of scytonemin, an ultraviolet

sunscreen pigment from the sheaths of cyanobacteria. *Experientia* 49:825–829.

Rasmussen, R.A. and Khalil, M.A.K. (1981) Atmospheric methane ($CH_4$): trends and seasonal cycles. *J Geophys Res* 86, doi:10.1029/JC086iC10p09826.

Robinson, T.D., Meadows, V.S., and Crisp, D. (2010) Detecting oceans on extrasolar planets using the glint effect. *Astrophys J* 721:L67–L71.

Robinson, T.D., Meadows, V.S., Crisp, D., Deming, D., A'Hearn, M.F., Charbonneau, D., Livengood, T.A., Seager, S., Barry, R.K., Hearty, T., Hewagama, T., Lisse, C.M., McFadden, L.A., and Wellnitz, D.D. (2011) Earth as an extrasolar planet: Earth model validation using EPOXI Earth observations. *Astrobiology* 11:393–408.

Robinson, T.D., Ennico, K., Meadows, V.S., Sparks, W., Bussey, D.B.J., Schwieterman, E.W., and Breiner, J. (2014) Detection of ocean glint and ozone absorption using LCROSS Earth observations. *Astrophys J* 787, doi:10.1088/0004-637X/787/2/171.

Rothman, L.S., Gordon, I.E., Barbe, A., Benner, D.C., Bernath, P.F., Birk, M., Boudon, V., Brown, L.R., Campargue, A., Champion, J.-P., Chance, K., Coudert, L.H., Dana, V., Devi, V.M., Fally, S., Flaud, J.-M., Gamache, R.R., Goldman, A., Jacquemart, D., Kleiner, I., Lacome, N., Lafferty, W.J., Mandin, J.-Y., Massie, S.T., Mikhailenko, S.N., Miller, C.E., Moazzen-Ahmadi, N., Naumenko, O.V., Nikitin, A.V., Orphal, J., Perevalov, V.I., Perrin, A., Predoi-Cross, A., Rinsland, C.P., Rotger, M., Simeckova, M., Smith, M.A.H., Sung, K., Tashkun, S.A., Tennyson, J., Toth, R.A., Vandaele, A.C., and Vander Auwera, J. (2009) The HITRAN 2008 molecular spectroscopic database. *J Quant Spectrosc Radiat Transf* 110:533–572.

Sagan, C., Thompson, W.R., Carlson, R., Gurnett, D., and Hord, C. (1993) A search for life on Earth from the Galileo spacecraft. *Nature* 365:715–721.

Saito, T., Terato, H., and Yamamoto, O. (1994) Pigments of *Rubrobacter radiotolerans. Arch Microbiol* 162:414–421.

Saito, T., Miyabe, Y., Ide, H., and Yamamoto, O. (1997) Hydroxyl radical scavenging ability of bacterioruberin. *Radiation Physics and Chemistry* 50:267–269.

Sanromá, E., Pallé, E., and García Munõz, A. (2013) On the effects of the evolution of microbial mats and land plants on the Earth as a planet. Photometric and spectroscopic light curves of paleo-Earths. *Astrophys J* 766, doi:10.1088/0004-637X/766/2/133.

Sanromá, E., Pallé, E., Parenteau, M.N., Kiang, N.Y., Gutiérrez-Navarro, A.M., López, R., and Montañés-Rodríguez, P. (2014) Characterizing the purple Earth: modeling the globally integrated spectral variability of the Archean Earth. *Astrophys J* 780, doi:10.1088/0004-637X/780/1/52.

Schabereiter-Gurtner, C., Piñar, G., Vybiral, D., Lubitz, W., and Rölleke, S. (2001) *Rubrobacter*-related bacteria associated with rosy discolouration of masonry and lime wall paintings. *Arch Microbiol* 176:347–354.

Schloss, P.D., Allen, H.K., Klimowicz, A.K., Mlot, C., Gross, J.A., Savengsuksa, S., McEllin, J., Clardy, J., Ruess, R.W., and Handelsman, J. (2010) Psychrotrophic strain of *Janthinobacterium lividum* from a cold Alaskan soil produces prodigiosin. *DNA Cell Biol* 29:533–541.

Schopf, J.W. (1993) Microfossils of the early Archean Apex chert: new evidence of the antiquity of life. *Science* 260:640–646.

Schuerger, A.C., Ulrich, R., Berry, B.J., and Nicholson, W.L. (2013) Growth of *Serratia liquefaciens* under 7 mbar, 0°C, and $CO_2$-enriched anoxic atmospheres. *Astrobiology* 13:115–131.

Seager, S., Turner, E.L., Schafer, J., and Ford, E.B. (2005) Vegetation's red edge: a possible spectroscopic biosignature of extraterrestrial plants. *Astrobiology* 5:372–390.

Seager, S., Schrenk, M., and Bains, W. (2012) An astrophysical view of Earth-based metabolic biosignature gases. *Astrobiology* 12:61–82.

Seager, S., Cash, W.C., Kasdin, N.J., Sparks, W.B., Turnbull, M.C., Kuchner, M.J., Roberge, A., Domagal-Goldman, S., Shaklan, S., Thomson, M., Lisman, D., Martin, S., Cady, E., and Webb, D. (2014) Exo-S: a probe-scale space mission to directly image and spectroscopically characterize exoplanetary systems using a starshade and telescope system [abstract 311.06]. In *American Astronomical Society Meeting #224*, American Astronomical Society, Washington, DC.

Segura, A., Kasting, J.F., Meadows, V., Cohen, M., Scalo, J., Crisp, D., Butler, R.A.H., and Tinetti, G. (2005) Biosignatures from Earth-like planets around M dwarfs. *Astrobiology* 5:706–725.

Shahmohammadi, H.R., Asgarani, E., Terato, H., Saito, T., Ohyama, Y., Gekko, K., Yamamoto, O., and Ide, H. (1998) Protective roles of bacterioruberin and intracellular KCl in the resistance of *Halobacterium salinarium* against DNA-damaging agents. *J Radiat Res* 39:251–262.

Solovchenko, A.E. and Merzlyak, M.N. (2008) Screening of visible and UV radiation as a photoprotective mechanism in plants. *Russ J Plant Physiol* 55:719–737.

Stapelfeldt, K.R., Brenner, M.P., Warfield, K., Belikov, R., Brugarolas, P., Bryden, G., Cahoy, K.L., Chakrabarti, S., Dekens, F., Effinger, R., Hirsch, B., Kissil, A., Krist, J.E., Lang, J., Marley, M.S., McElwain, M.W., Meadows, V., Nissen, J., Oseas, J., Serabyn, G., Sunada, E., Traub, W.A., Trauger, J.T., and Unwin, S.C. (2014) Exo-C: a probe-scale space mission to directly image and spectroscopically characterize exoplanetary systems using an internal coronagraph [abstract #311.07]. In *American Astronomical Society Meeting #224*, American Astronomical Society, Washington, DC.

Takano, H., Asker, D., Beppu, T., and Ueda, K. (2006) Genetic control for light-induced carotenoid production in non-phototrophic bacteria. *J Ind Microbiol Biotechnol* 33:88–93.

Tanaka, Y., Sasaki, N., and Ohmiya, A. (2008) Biosynthesis of plant pigments: anthocyanins, betalains and carotenoids. *Plant J* 54:733–749.

Tian, B., Sun, Z., Xu, Z., Shen, S., Wang, H., and Hua, Y. (2008) Carotenoid 3',4'-desaturase is involved in carotenoid biosynthesis in the radioresistant bacterium *Deinococcus radiodurans*. *Microbiology* 154:3697–3706.

Tian, F., France, K., Linsky, J.L., Mauas, P.J.D., and Vieytes, M.C. (2014) High stellar FUV/NUV ratio and oxygen contents in the atmospheres of potentially habitable planets. *Earth Planet Sci Lett* 385:22–27.

Tinetti, G., Meadows, V.S., Crisp, D., Fong, W., Fishbein, E., Turnbull, M., and Bibring, J.-P. (2006a) Detectability of planetary characteristics in disk-averaged spectra. I: the Earth model. *Astrobiology* 6:34–47.

Tinetti, G., Meadows, V.S., Crisp, D., Kiang, N.Y., Kahn, B.H., Bosc, E., Fishbein, E., Velusamy, T., and Turnbull, M.

(2006b) Detectability of planetary characteristics in disk-averaged spectra II: synthetic spectra and light-curves of Earth. *Astrobiology* 6:881–900.

Tinetti, G., Rashby, S., and Yung, Y.L. (2006c) Detectability of red-edge-shifted vegetation on terrestrial planets orbiting M stars. *Astrophys J* 644:L129–L132.

Ungers, G.E. and Cooney, J.J. (1968) Isolation and characterization of carotenoid pigments of *Micrococcus roseus*. *J Bacteriol* 96:234–241.

Valadon, L.R. and Mummery, R.S. (1968) Carotenoids in floral parts of a narcissus, a daffodil and a tulip. *Biochemical J* 106:479–484.

Venil, C. and Lakshmanaperumalsamy, P. (2009) An insightful overview on microbial pigment, prodigiosin. *Electronic Journal of Biology* 5:49–61.

Wahlund, T.M., Woese, C.R., Castenholz, R.W., and Madigan, M.T. (1991) A thermophilic green sulfur bacterium from New Zealand hot springs, *Chlorobium tepidum* sp. nov. *Arch Microbiol* 156:81–90.

Wiggli, M., Smallcombe, A., and Bachofen, R. (1999) Reflectance spectroscopy and laser confocal microscopy as tools in an ecophysiological study of microbial mats in an alpine bog pond. *J Microbiol Methods* 34:173–182.

Williams, D.M. and Gaidos, E. (2008) Detecting the glint of starlight on the oceans of distant planets. *Icarus* 195:927–937.

Williams, P., Winzer, K., Chan, W.C., and Cámara, M. (2007) Look who's talking: communication and quorum sensing in the bacterial world. *Philos Trans R Soc Lond B Biol Sci* 362:1119–1134.

Williams, W.E., Gorton, H.L., and Vogelmann, T.C. (2003) Surface gas-exchange processes of snow algae. *Proc Natl Acad Sci USA* 100:562–566.

Ziegelhoffer, E.C. and Donohue, T.J., (2009) Bacterial responses to photo-oxidative stress. *Nat Rev Microbiol* 7:856–863.

Address correspondence to:
*Edward W. Schwieterman*
*Box 351580*
*University of Washington*
*Seattle, WA 98115-1580*

*E-mail:* eschwiet@astro.washington.edu

**Abbreviations Used**

DSMZ = Deutsche Sammlung von Mikroorganismen und Zellkulturen
VRE = vegetation red edge